\documentclass[useAMS]{mn2e}
\usepackage{times}
\usepackage{amssymb}
\usepackage{amsmath}
\usepackage{rotating}
\usepackage{epsfig}
\input{epsf}

\title[What Powers the Relativistic Jets in FSRQs?]
{What Powers the Most Relativistic Jets? II: Flat Spectrum Radio Quasars}

\author[E. Gardner and C. Done]
{Emma Gardner and Chris Done\\
Centre for Extragalactic Astronomy, Department of Physics, University of Durham, South Road,
Durham DH1 3LE, UK\\}

\date{Submitted to MNRAS}
\pagerange{\pageref{firstpage}--\pageref{lastpage}} \pubyear{}

\def\apj{ApJ}

\begin{document}

\topmargin = -0.5cm

\maketitle

\label{firstpage}

\begin{abstract}

  Flat Spectrum Radio Quasars (FSRQs) are the most powerful
  relativistic jets seen from supermassive black holes (BHs) accreting via a
  radiatively efficient thin disc. Their high energy emission is well
  modelled by highly relativistic electrons in the jet Compton
  upscattering an external source of seed photons, primarily from the
  broad line region. Strong Doppler boosting by the jet bulk motion
  makes these FSRQs readily detectable by the Fermi Large Area
  Telescope. We combine jet spectral models with scaling relations for the
  jet physical parameters as a function of mass and accretion
  rate. This does not match well to the Gamma-ray loud Narrow Line
  Seyfert 1s, assuming their low BH masses are
  reliable, but is able to predict much of the spectral evolution
  observed along the Blazar sequence. We use these models in
  conjunction with cosmological simulations of efficiently accreting
  BH number densities, and find that they overpredict the observed
  number of FSRQs by 2 orders of magnitude if all of these objects
  produce a FSRQ jet. We can better reproduce the observed numbers if
  jets are only produced by high spin BHs and BH spin is built from
  chaotically aligned accretion episodes so that high spin BHs are
  rare. However, this does not reproduce the observed redshift and
  mass accretion rate distributions of the FSRQs. This may indicate a
  redshift dependence in accretion mode, with sustained alignment accretion
  episodes being more prevalent at higher redshift, or that
  there is some other trigger for FSRQ jets.

\end{abstract}


\section{Introduction}

Blazars are the most extreme examples of relativistic jets from
supermassive black holes (BHs). They represent a class of active
galactic nuclei (AGN) where the jet is highly relativistic
($\Gamma>10$) and closely aligned with our line of sight. The jet
emission is strongly Doppler boosted by the relativistic bulk motion
and consequently dominates the spectrum of the AGN, from radio up to
gamma-rays. As a result these are the most numerous sources detected
by the Fermi/LAT satellite in the GeV regime (Nolan et al. 2012).

  Blazars can be divided into two types -- BL Lacs and flat spectrum
  radio quasars (FSRQs). This division is made on an arbitrary upper
  limit to the observed equivalent width of the emission lines, but
  this typically correlates with the broad band continuum properties,
  where the jet spectra of BL Lacs consist of a low energy hump from
  synchrotron emission and a second, similar luminosity higher energy
  hump from synchrotron self-Compton (SSC) emission.  The spectra of
  FSRQ jets also show a low energy peak from synchrotron emission but
  have a much more luminous Compton hump, often by more than an order
  of magnitude (Fossati et al. 1998). This can be explained by there
  being an intrinsic difference between the majority of BL Lacs and
  FSRQs (a small fraction of objects are  misclassified due to selection effects:
  Giommi et al. 2012), such that
  the BL Lacs do not have additional seed photons from outside of the
  jet. In contrast, the FSRQs have a disc and associated broad line region
  (BLR), which provide an additional external source of seed photons for Compton
  scattering, leading to the observed Compton peak dominance in FSRQs
  (e.g. Ghisellini et al. 2017).

This change in seed photons plausibly occus because of a change in the
accretion mode similar to that seen in the black hole binary systems
(BHBs; see e.g. Done, Gierlinski \& Kubota 2007). FSRQs are high
accretion rate AGN containing highly accreting black holes
($\dot{m}=\dot{M}/\dot{M}_{Edd}>0.01$, where
$\eta\dot{M}_{Edd}c^2=L_{Edd}$). At high accretion rates, the
accretion flow around the BH takes the form of a radiatively efficient
accretion disc, which can often be seen dominating in the optical-UV
in FSRQ spectra, despite the strong jet emission (Ghisellini et
al. 2010, hereafter G10). The strong UV disc emission illuminates
material above the disc, which at a particular radius, set by the gas
density and illuminating flux, re-emits the radiation in the form of
broad emission lines (the `broad line region'). Some fraction of
the accretion disc emission is also reprocessed by the
torus. Together, the UV bright accretion disc, BLR and reprocessed
emission from the torus all act as sources of external seed photons
for the jet. However, crucially, the radius at which the broad lines
are produced is normally at a larger distance from the BH than the
region of the jet where the high energy emission is produced
(Ghisellini \& Tavecchio 2009). As a result, the jet electrons moving
with the relativistic bulk motion of the jet see the stationary BLR
seed photons strongly Doppler boosted and this greatly enhances the
external Compton emission of the FSRQ.

In contrast, the BHs responsible for the production of BL Lac jets are
at much lower accretion rates ($\dot{m}<0.01$), where the accretion
flow switches from a geometrically thin, radiatively efficient UV
bright disc to a hotter, geometrically thick, radiatively inefficient
flow (e.g Advection Dominated Accretion Flow: ADAF Narayan \& Yi
1995). The switch to a radiatively inefficient accretion flow means
there is no UV bright inner disc to illuminate the BLR and provide
external seed photons. As a result, BL Lac spectra lack both disc
emission and broad lines (Stickel et al. 1991) and their high energy
Compton humps include only SSC of synchrotron emission generated
intrinsically within the jet (Ghisellini, Maraschi \& Tavecchio 2009). For the remainder of this paper, we assume all objects with $\dot{m}\ge
0.01$ have a disc/BLR/torus and refer to these as FSRQs, while those
with $\dot{m}<0.01$ have none of these external components and we refer
to these as BL Lacs.

Whilst the mechanisms by which blazar jets emit radiation are
relatively well understood, where the energy comes from to power these
jets in the first place is not. These jets have bulk Lorentz factors
of 10-15 and estimates of the jet power in FSRQs put it at the order
of the accretion power or above (G10, Ghisellini et al. 2014). This
requires tapping the spin energy of the BH (Blandford \& Znajek 1977)
and means that the most relativistic jets should necessarily be
produced by the most highly spinning BHs.

Gardner \& Done (2014, hereafter Paper 1) took a
statistical approach to this problem. Rather than studying individual
sources, Paper 1 concentrated on modelling the population of Fermi
blazars as a whole -- specifically the population of BL Lacs, since
they do not have the added complication of external seed photon
sources. Cosmological simulations predict the number of BHs accreting
at each redshift as a function of mass and accretion rate. Since BL
Lac jets are produced when $\dot{m}<0.01$, they should only be produced
by low accretion rate BHs. Paper 1 initially assumed that all BHs with
$\dot{m}<0.01$ produced a BL Lac type jet. The predicted numbers of
these BHs were taken from the Millennium Simulation (Springel et al.
2005; Fanidakis et al. 2011; 2012) and each BH was then assigned a jet
spectrum, appropriately scaled to its mass and accretion rate,
assuming all size scales in the jet scale with BH mass and the power
in particles and jet magnetic fields is a fixed fraction of the
accretion power. Each jet was then given a random orientation and its
resulting redshifted flux calculated to determine whether it would be
bright enough to be detected by Fermi. This predicted population of
Fermi detected BL Lacs was found to overpredict the observed number of
Fermi detected BL Lacs by 3 orders of magnitude. Producing a BL Lac
jet requires a BH with $\dot{m}<0.01$, but clearly not every BH with
$\dot{m}<0.01$ produces a BL Lac jet (see e.g. Wilson \& Colbert 1995;
Moderski, Sikora \& Lasota 1998; Padovani et al. 2015).

Paper 1 found that the observed number of BL Lacs was much better
reproduced if production of BL Lac jets was restricted to BHs with
$\dot{m}<0.01$ and high spin ($a>0.8$). Maraschi et al. (2012) argue
that the efficiency of spin-powered jet production drops off sharply
below $0.8$, so that this forms an effective spin threshold for
relativistic jet production. This reduces the predicted population
sufficiently if high spin BHs are rare. Aligned accretion is very
efficient at spinning up the BH (e.g. Volonteri et al. 2007). However,
randomly aligned small accretion episodes (chaotic accretion) result
in low spin (King et al. 2008), with high spin BHs produced only through
BH-BH mergers (Fanidakis et al. 2011; 2012). Restricting production of
BL Lac jets to high spin, low accretion rate BHs and assuming chaotic
accretion not only allows the total number of Fermi detected BL Lacs
to be reproduced, but also better matches the observed mass and
redshift distributions. This is because the low accretion rate, high
spin BHs are those that are formed latest in gas-poor mergers which
produce the most massive BHs. This suggests that high spin may be
required to produce the highly relativistic jets in BL Lacs.

In this paper, we use the same method to try to predict the observed
population of Fermi detected FSRQs. The scaling relations are able to
reproduce the FSRQ blazar sequence with increasing BH mass (Ghisellini
et al. 2017), but the small population of gamma-ray loud NLS1s
($\gamma$NLS1s) are more Compton dominant than standard jet scaling
relations predict. This suggests that, either FSRQ jets do not follow
standard jet scaling relations, or $\gamma$NLS1 masses may be larger
than previously estimated, as has been suggested by Calderone et
al. (2013), Baldi et al. (2016) and D'Ammando et al. (2017).

We again use the BH number
densities predicted by the Millennium Simulation and this time assume
that all BHs accreting with $\dot{m}>0.01$ produce a FSRQ type jet. We
extend the spectral model of Paper 1 to include external sources of
seed photons and assume the same standard jet scalings (sizescales
scale with $M_{BH}$ and power in particles and magnetic fields scales
with accretion power), which mimic the observed spectral changes in BL
Lacs (Paper 1). As in Paper 1, we find that assuming all BHs in the
appropriate accretion regime produce a highly relativistic jet
overpredicts the observed FSRQ population, however by not as much
(only 2 orders of magnitude, rather than 3 in the case of BL
Lacs). Again, we try imposing a spin cut, such that only high spin,
high accretion rate BHs produce FSRQ jets. However, although this
allows us to better match the observed number of Fermi-detected FSRQs,
we find we cannot match the observed mass, accretion rate or redshift
distributions -- particularly the tail out to high redshifts
($z>2$). This is due to a lack of high mass, high spin BHs in the
cosmological simulations at redshift $2-3$. If production of a FSRQ jet
really does require a high spin BH, our simulations suggest there
should be more high mass, high spin BHs at high redshift than a solely
chaotic accretion model predicts. This suggests there may be a trend
from chaotic accretion towards more prolonged accretion (which spins
up rather than spins down the BH) at higher redshifts, as also
suggested by Dotti et al. (2013) and Dubois et al. (2014).

\section{External-Compton Jets}

We extend the single-zone SSC model of Paper 1 to include sources of
external seed photons, since these are important in the higher
accretion rate FSRQs, which have UV bright accretion discs. We code
up the model of Ghisellini \& Tavecchio (2009), including seed photons
from the accretion disc and X-ray corona, emission from the BLR and
torus, and reflection of coronal X-rays off the BLR.  We follow their
notation below, where quantities in the jet frame are primed when
there could be confusion, but not where it is self-evident, e.g. jet
quantities such as electron Lorentz factors and jet magnetic field.

We have made this code publicly available within the {\sc{xspec}}
spectral fitting package. We briefly summarise the model here, with
full details in the Appendix.

We assume a spherical emission region of radius $R_{diss}$ and a
conical jet, such that $R_{diss}$ is related to the distance of the
emission region from the black hole by $Z_{diss}=\phi R_{diss}$, where
$\phi$ is the half opening angle of the jet. We assume the jet
emission is dominated by this single spherical emission region at the
jet base and neglect the contribution from regions further out along
the jet, as this mostly affects the low energy (predominantly radio)
emission. We assume material in the jet moves at a constant bulk
Lorentz factor ($\Gamma$), and that a fraction of the resulting jet
power is used to accelerate electrons in the emission region. The
power injected into relativistic electrons in the jet frame is then $P^\prime_{rel}=4/3\pi
R_{diss}^3 \int\gamma m_ec^2 Q(\gamma) d\gamma$, where the accelerated
electron distribution is a broken power law of the form:

\begin{multline}
Q(\gamma)=Q_0 \frac {\left(\frac{\gamma}{\gamma_b}\right)^{-s_1} }
{ \left[ 1+\left(\frac{\gamma}{\gamma_b}\right)^{-s_1+s_2}\right] } \\
\mbox{ for } \gamma_{min}<\gamma<\gamma_{max}
\end{multline}
 
We note that this is slightly different from the form used in Paper
1. Firstly we now use $s$ rather than $n$ to denote in the injected
power law indices in order to be consistent with Ghisellini \&
Tavecchio (2009). Secondly, Paper 1 has the denominator as
$(1-\gamma/\gamma_b)^{-s_1+s_2}$ rather than the form used here and in
Ghisellini \& Tavecchio (2009). We have run tests and find that this
change generally gives less than a factor 30 per cent difference in
the resultant spectra. 

These electrons cool by emitting self-absorbed synchrotron and
synchrotron self-Compton radiation and by upscattering seed photons
from external sources of radiation. We assume the distance of the BLR
and infra-red torus ($R_{BLR}$ and $R_{IR}$) from the central
black hole scale with the accretion disc luminosity as (Ghisellini \&
Tavecchio 2009):

\begin{equation}
R_{BLR} = 10^{17}(\frac{L_d}{10^{45} erg s^{-1}})^{1/2} cm 
\end{equation}
\begin{equation}
R_{IR} = 2.5\times10^{18}(\frac{L_d}{10^{45} erg s^{-1}})^{1/2} cm
\end{equation}

The total seed photon energy density in the jet frame is $U^\prime_{seed}=U^\prime_B+g(\gamma)(U^\prime_{sync}+U^\prime_{ex})$ and therefore includes both the magnetic energy
density, $U^\prime_B=B^2/8\pi$, and the fraction $g(\gamma)$ of the energy
density of synchrotron, $U^\prime_{sync}$, and external, $U^\prime_{ex}$, seed photons which can be Compton
upscattered by electrons of energy $\gamma$ within the Klein-Nishina
limit. The accelerated electron distribution cools into a steady state electron distribution,
$N(\gamma)=-\dot{\gamma}^{-1}\int_\gamma^{\gamma_{max}} Q(\gamma_i)d\gamma_i$, where 
the rate at which an electron loses energy is $\dot{\gamma}
m_ec^2=4/3\gamma^2 \sigma_T c U^\prime_{seed}$. Since the
cooling timescale $t_{cool}=\gamma/\dot{\gamma}$ depends on $\gamma$, with high energy electrons cooling fastest, we
calculate the electron Lorentz factor that can just cool in a light crossing
time of the region,$\gamma_{cool}$, and join smoothly onto the
accelerated uncooled electron distribution below this. The full self-consistent
electron distribution is then characterised by $N(\gamma)=Kn(\gamma)$,
where $K$ is the number density of electrons at $\gamma=1$ and
$n(\gamma)$ incorporates all the spectral shape.
We calculate the resulting (self-absorbed) synchrotron and Compton emission using the delta function approximation as this is
much faster than using the full kernel but is accurate enough for our
statistical analysis (Dermer \& Menon 2009).

This jet frame emission is boosted by the bulk motion of the jet,
where the Doppler factor of the boosting ($\delta$) depends on both
$\Gamma$ and the orientation of the jet as
$\delta=(\Gamma-\cos{\theta}\sqrt{\Gamma^2-1})^{-1}$ and we transform
jet frame frequencies ($\nu^\prime$) and luminosities ($L^\prime$) to
observed frame quantities as: $\nu=\delta\nu^\prime$ and
$L=\delta^3L^\prime$. Dermer \& Menon 2009 advocate multiplying by an
additional factor of $\delta$ for the special case of a seed photon
source that is ahead of the jet emission region, to account for the
anisotropy in the seed photon field in the jet frame. However, an
anisotropic seed photon field produces anisotropic cooling and
therefore an anisotropic electron distribution, which can effectively
cancel out the effects of the seed photon anisotropy (Ghisellini et
al. 1989; Ghisellini et al. 1991; Gierli{\'n}ski et al. 1999), assuming that the electrons are not re-isotropised by turbulent
scattering on timescales much shorter than the cooling time. Since
we include cooling from multiple seed photons sources (accretion flow,
BLR and torus), with multiple orientations with respect to the jet
producing multiple anisotropies, we choose to adopt
$\nu=\delta\nu^\prime$ and $L=\delta^3L^\prime$ in all cases.

Finally, the jet emission is  
cosmologically redshifted and attenuated due to pair production on the
extragalactic infra-red background light (though this is generally small
for the Fermi bandpass) to produce the observed flux. 

The parameters of our model are therefore: 
\begin{itemize}
\item Parameters of the accretion flow: black hole mass and Eddington scaled accretion rate ($M_{BH}$ and $\dot{m}$), for calculating the density of external seed photons. 
\item
Physical parameters of the jet: $\Gamma$, radius of emission region ($R_{diss}$) and half opening angle of the jet ($\phi$).
\item
The magnetic field of the emission region and power injected into relativistic electrons ($B$ and $P^\prime_{rel}$).
\item
Parameters of the injected electron distribution: $\gamma_{min}$, $\gamma_b$, $\gamma_{max}$, $s_1$ and $s_2$.
\end{itemize}

We adopt the cosmology used in the Millennium simulations: $h=0.72,$ $\Omega_m=0.25$, $\Omega_{vac}=0.75$ (Springel et al. 2005; Fanidakis et al. 2011).

\section{Scaling Jets}\label{sec:scalings}

As in Paper 1, we assume that the acceleration mechanism is the same
for all FSRQs, giving the same injected electron distribution,
regardless of mass and accretion rate. We also assume all jets are
produced with the same $\Gamma$ and the same half opening angle (which
we fix to $\phi=0.1$). This leaves three remaining parameters:
$R_{diss}$, $B$ and $P^\prime_{rel}$.

We assume that FSRQs follow the same standard jet scalings that BL
Lacs appear to follow (Paper 1). Hence we scale $R_{diss}\propto M$,
since all size scales should scale with the mass of the black hole
(Heinz \& Sunyaev 2003), and assume the jet power is a constant
fraction of the total accretion power, $P_j\propto\dot{m}M$. We again
stress that this assumption is valid whether the jet is powered by the
accretion flow or the spin energy of the black hole, since extraction
of black hole spin energy relies on magnetic fields generated in the
accretion flow, which will be affected by accretion rate. A constant
fraction of the total jet power is then injected into relativistic
particles and magnetic fields. Hence $P^\prime_{rel}\propto
P_j\propto\dot{m}M$ and $B\propto U^{\prime 1/2}_B\propto
(P^\prime_B/R_{diss}^2)^{1/2} \propto (P_j/R_{diss}^2)^{1/2} \propto
(\dot{m}/M)^{1/2}$ (Paper 1).

We choose the mean FSRQ parameters from G10 to scale from, which are
the logarithmic average values from their sample of 53 Fermi detected
FSRQs. This gives $M_0 = 1\times10^9\,M_\odot$, $\dot{m}=0.1$, $R_0 =
1.89\times10^{16} cm$, $B_0 = 2.6G$, $P^\prime_{rel,0}=2\times10^{43} erg
s^{-1}$, $\Gamma=13$, $\gamma_{min}=1$, $\gamma_b=300$,
$\gamma_{max}=3\times10^3$, $s_1=1$, $s_2=2.7$, and we scale
$R_{diss}$, $P^\prime_{rel}$ and $B$ as:

\begin{equation}
R_{diss} = R_0\frac{M}{M_0}
\end{equation}
\begin{equation}
P^\prime_{rel} = P^\prime_{rel,0}\frac{\dot{m}}{\dot{m_0}}\frac{M}{M_0} 
\end{equation}
\begin{equation}
B = B_0\left(\frac{\dot{m}}{\dot{m_0}}\frac{M_0}{M}\right)^{1/2}
\end{equation}

The distance to the BLR and IR torus are both $\propto
L_d^{1/2}$. This implies $R_{BLR}$ and $R_{IR}$ should also scale with
the mass and accretion rate of the black hole, since $L_d \propto
\dot{m}M$. Hence $R_{BLR}\propto R_{IR} \propto (\dot{m}M)^{1/2}$.

\subsection{Spectral Changes with Mass}

\begin{figure} 
\centering
\begin{tabular}{l}
\leavevmode  
\epsfxsize=8cm \epsfbox{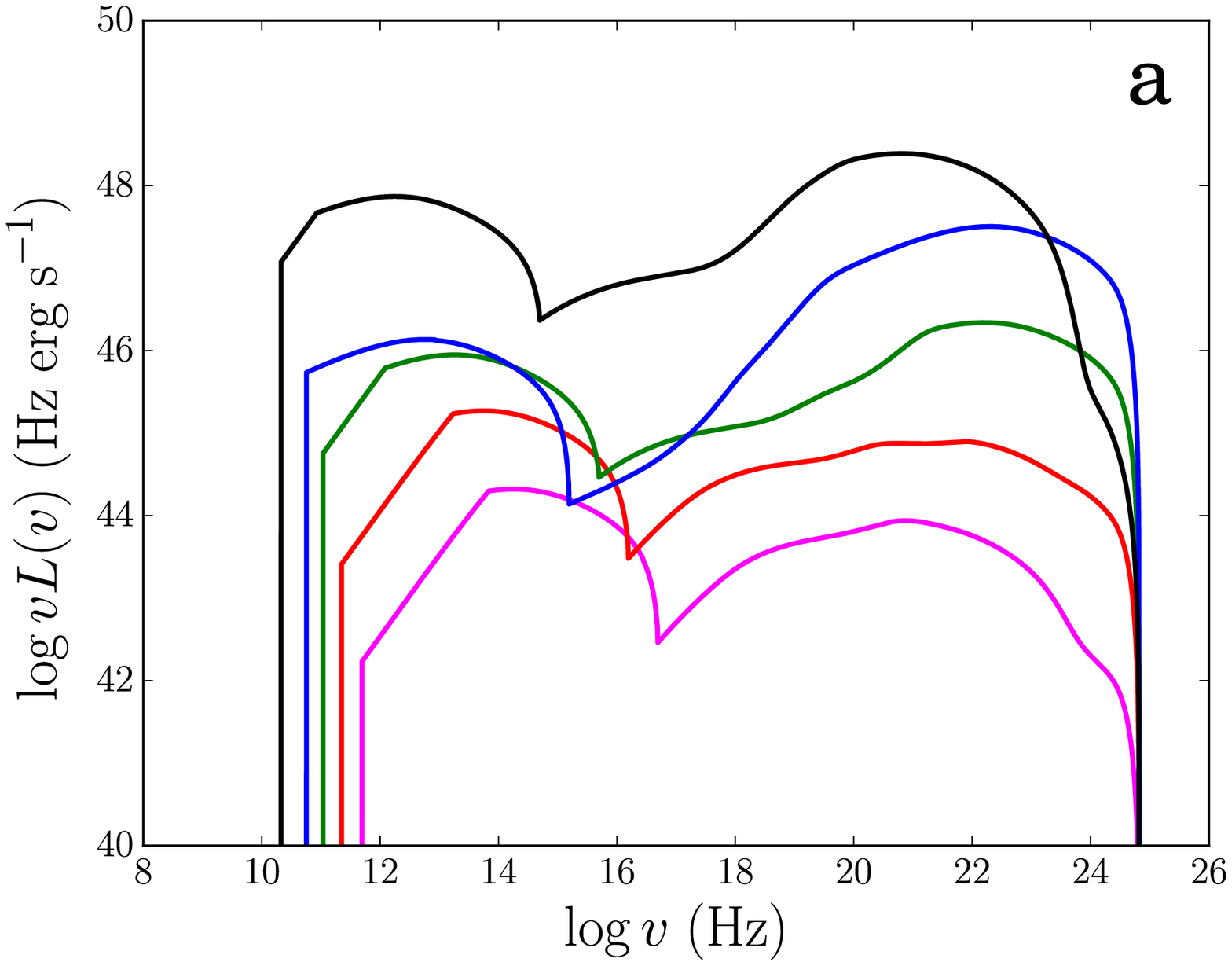} \\
\epsfxsize=8cm \epsfbox{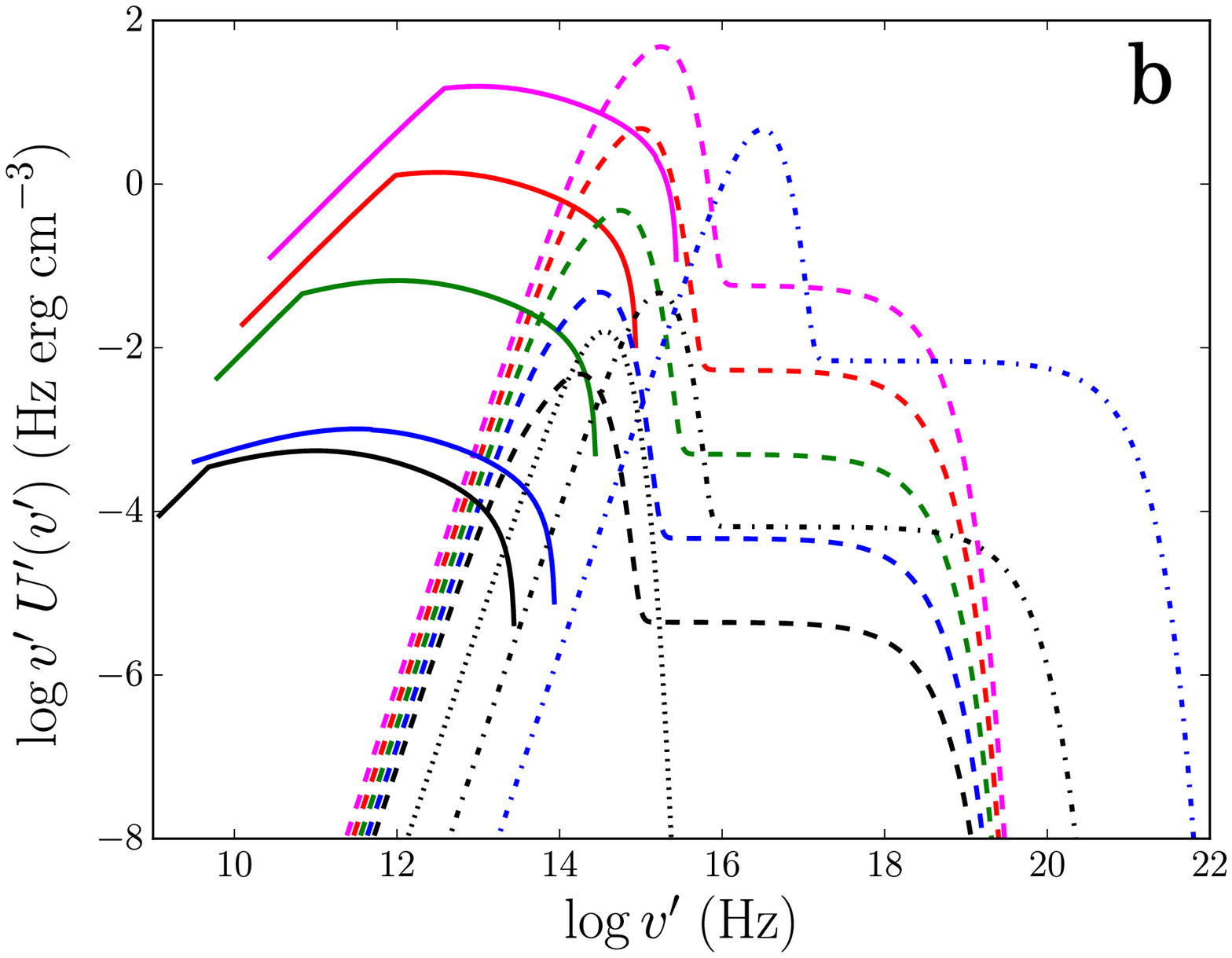} \\
\epsfxsize=8cm \epsfbox{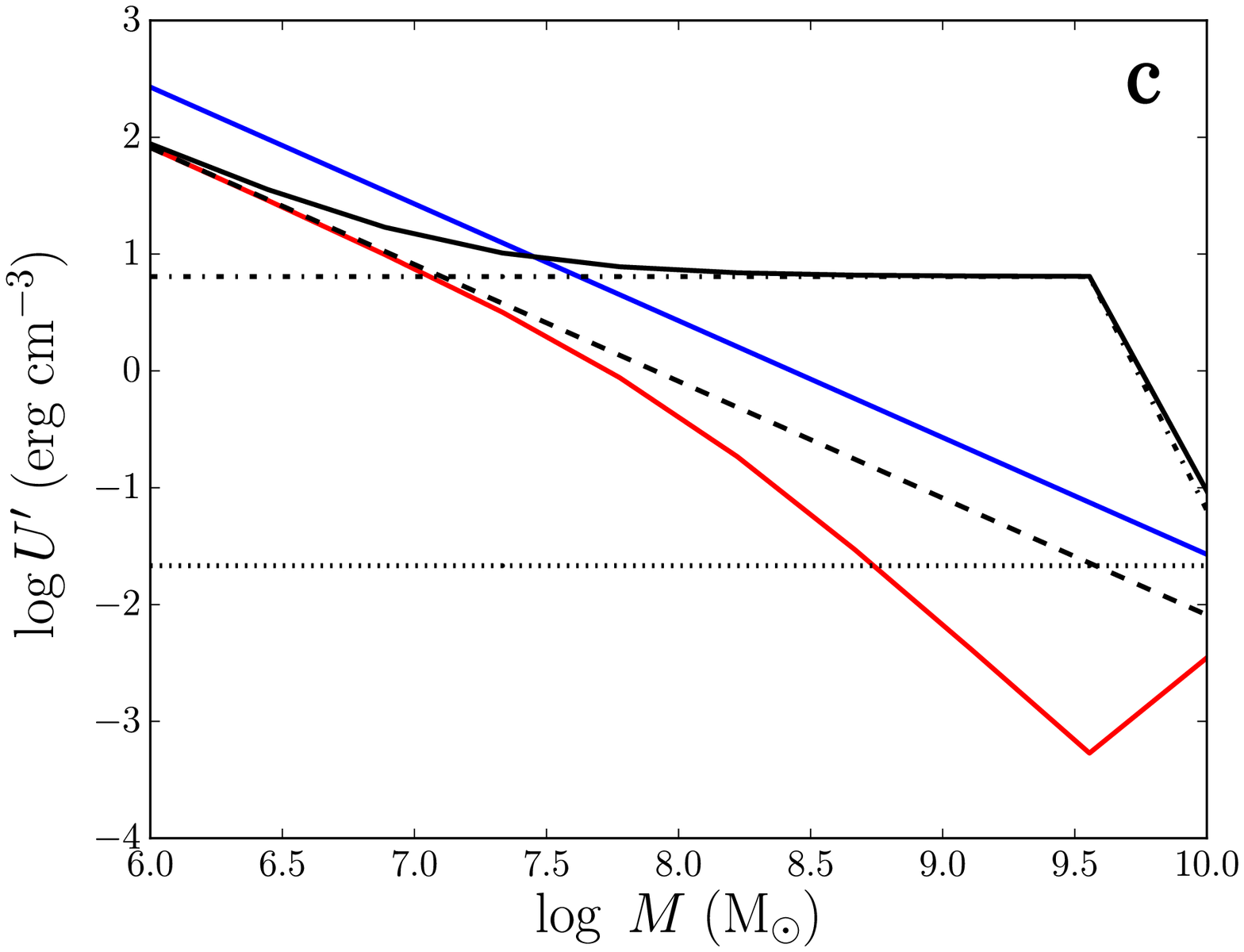} \\
\end{tabular}
\caption{Spectral changes with mass using standard jet scalings ($R_{diss}\propto M$, $P^\prime_{rel}\propto \dot{m}M$, $B\propto (\dot{m}/M)^{1/2}$). a). FSRQ model SEDs for fixed accretion rate and increasing BH mass ($M_{BH}=10^{6}$ (magenta), $10^{7}$ (red), $10^{8}$ (green), $10^{9}$ (blue) and $10^{10} M_\odot$ (black), $\dot{m}=0.1$). b). Corresponding seed photon energy density spectra as seen in the jet frame. Solid lines show synchrotron seed photons, dashed lines show accretion disc plus coronal seed photons, dot-dashed lines show seed photons from the BLR plus coronal flux reflected by the BLR, and dotted lines show seed photons from the torus. The seed photon energy density from the torus is the same for all masses. The seed photon energy density from the BLR is the same for all masses except $10^{10} M_\odot$, where $Z_{diss}>R_{BLR}$. c). Seed photon energy densities in jet frame as a function of BH mass. Blue line shows $U^\prime_B$, red line shows $U^\prime_{sync}$, black lines show energy densities of external seed photons: dashed line shows $U^\prime_d+U^\prime_X$, dot-dashed line shows $U^\prime_{BLR}+U^\prime_{XBLR}$, dotted line shows $U^\prime_{IR}$, and solid line shows total $U^\prime_{ex}$.}
\label{fig1}
\end{figure}

Fig.\ref{fig1}a shows a sequence of FSRQ spectra with increasing BH mass. The accretion rate is fixed to $\dot{m}=0.1$ and $R_{diss}$, $B$ and $P^\prime_{rel}$ are scaled as described above. 

As BH mass decreases, so does the size of the emission region, since $R_{diss}\propto M$. This can be seen in the increase in synchrotron self-absorption frequency, from $\sim10^{10.5}$ (black spectrum) to $\sim10^{11.5}$ (magenta spectrum). The total luminosity also decreases, since the power injected into relativistic electrons decreases with decreasing mass ($P^\prime_{rel}\propto M$). 

The relative strengths of the synchrotron and Compton humps also changes with mass. The blue spectrum corresponds to the mean FSRQ model of G10, with $\dot{m}=0.1$, $M=10^{9}M_\odot$. It shows a strong Compton hump at $10^{22}$Hz due to Compton up-scattering of external seed photons, predominantly from the BLR. The low energy synchrotron hump is roughly an order of magnitude less luminous ($\sim10^{46}erg\,s^{-1}$). As the BH mass drops from $10^9$ (blue) - $10^6M_\odot$ (magenta), the relative luminosity of the Compton hump decreases until at the lowest masses the two humps show comparable luminosity. The relative strength of the two humps depends on the relative strength of the energy density in magnetic fields compared to the energy density of external seed photons. 

Fig.\ref{fig1}b shows the spectral energy density of seed photons in the jet frame. As mass drops so does the emission region size and hence its distance from the BH, since $Z_{diss}=R_{diss}/\phi\propto M$. Smaller $Z_{diss}$ increases the energy density of accretion disc seed photons, despite the drop in $L_d$ with $M$, showing an increase of $\sim3$ orders of magnitude (blue dashed line to magenta dashed line). However, the dominant source of seed photons is $U^\prime_{BLR}$ and this stays constant, since $U^\prime_{BLR}\propto L_d/R^2_{BLR}\propto L_d/(L_d^{1/2})^2=const$ for $Z_{diss}<R_{BLR}$ (blue dot-dashed line). In contrast the magnetic field, which determines the amount of synchrotron emission, increases as BH mass decreases, since $B\propto M^{-1/2}$. As a result, $U^\prime_{sync}$ (solid lines) increases by more than 4 orders of magnitude, becoming comparable to $U^\prime_{BLR}$ at the lowest masses. Consequently, the Compton humps of the lowest mass spectra are dominated by up-scattering of synchrotron radiation, causing them to look more like low accretion rate synchrotron self-Compton BL Lacs than FSRQs, despite their higher accretion rates. 

The lack of external seed photons means less efficient cooling in lower mass objects. For the $10^9M_\odot$ spectrum (Fig.\ref{fig1}a, blue line) the cooling is almost complete with $\gamma_{cool}=7$, where $\gamma_{cool}$ is the minimum Lorentz factor of electrons that can cool in one light crossing time. For the $10^6M_\odot$ spectrum (magenta), $\gamma_{cool}$ has increased to $10^6$, resulting in a clear spectral break at $\sim10^{13.5}$Hz in the synchrotron emission. The decreasing frequency of this spectral break tracks the decrease in $\gamma_{cool}$ and increase in cooling from $10^6$-$10^9M_\odot$. 

Above $10^9M_\odot$, $\gamma_{cool}$ increases again (Fig.\ref{fig1}a, black spectrum, $\gamma_{cool}=39$). This is because for a $10^{10}M_\odot$ BH the emission region has gone beyond $R_{BLR}$, since $Z_{diss}\propto M$ while $R_{BLR}$ (and $R_{IR}$) $\propto M^{1/2}$. This causes $U^\prime_{BLR}$ to drop dramatically (black dot-dashed line in Fig.\ref{fig1}c), reducing the amount of cooling. The next strongest source of seed photons is the torus (Fig.\ref{fig1}b, black dotted line). $U^\prime_{IR}$ is constant for all masses since like $U^\prime_{BLR}$, $U^\prime_{IR}\propto L_d/R^2_{IR}\propto L_d/(L_d^{1/2})^2=const$ for $Z_{diss}<R_{IR}$, which is the case for all five masses. Consequently, above $10^9M_\odot$ the ratio between synchrotron and Compton peaks drops again. 

Fig.\ref{fig1}c shows the total energy densities of seed photons in the jet frame as a function of BH mass. This shows clearly for masses around $10^9M_\odot$, where the energy density of BLR photons dominates (black dot-dashed line), the energy density of synchrotron radiation is suppressed (red line) due to the strong cooling. $U^\prime_{sync}$ recovers at higher masses as $Z_{diss}>R_{BLR}$ and dominates over $U^\prime_{BLR}$ at low masses ($<10^7M_\odot$) where the magnetic field is strongest (blue line). 

The sequence of spectra shown in Fig.\ref{fig1}a appear remarkably similar to the sequence of observed FSRQ spectra binned by luminosity shown by Ghisellini et al. 2017. Ghisellini et al. 2017 find that the Compton dominance of FSRQ spectra increases with luminosity, causing the X-ray spectral slope to harden. Fig.\ref{fig1}a shows that this can be explained if the higher luminosity bins are dominated by increasingly higher mass FSRQs.

\subsection{Spectral Changes with Accretion Rate}

\begin{figure} 
\centering
\begin{tabular}{l}
\leavevmode  
\epsfxsize=8cm \epsfbox{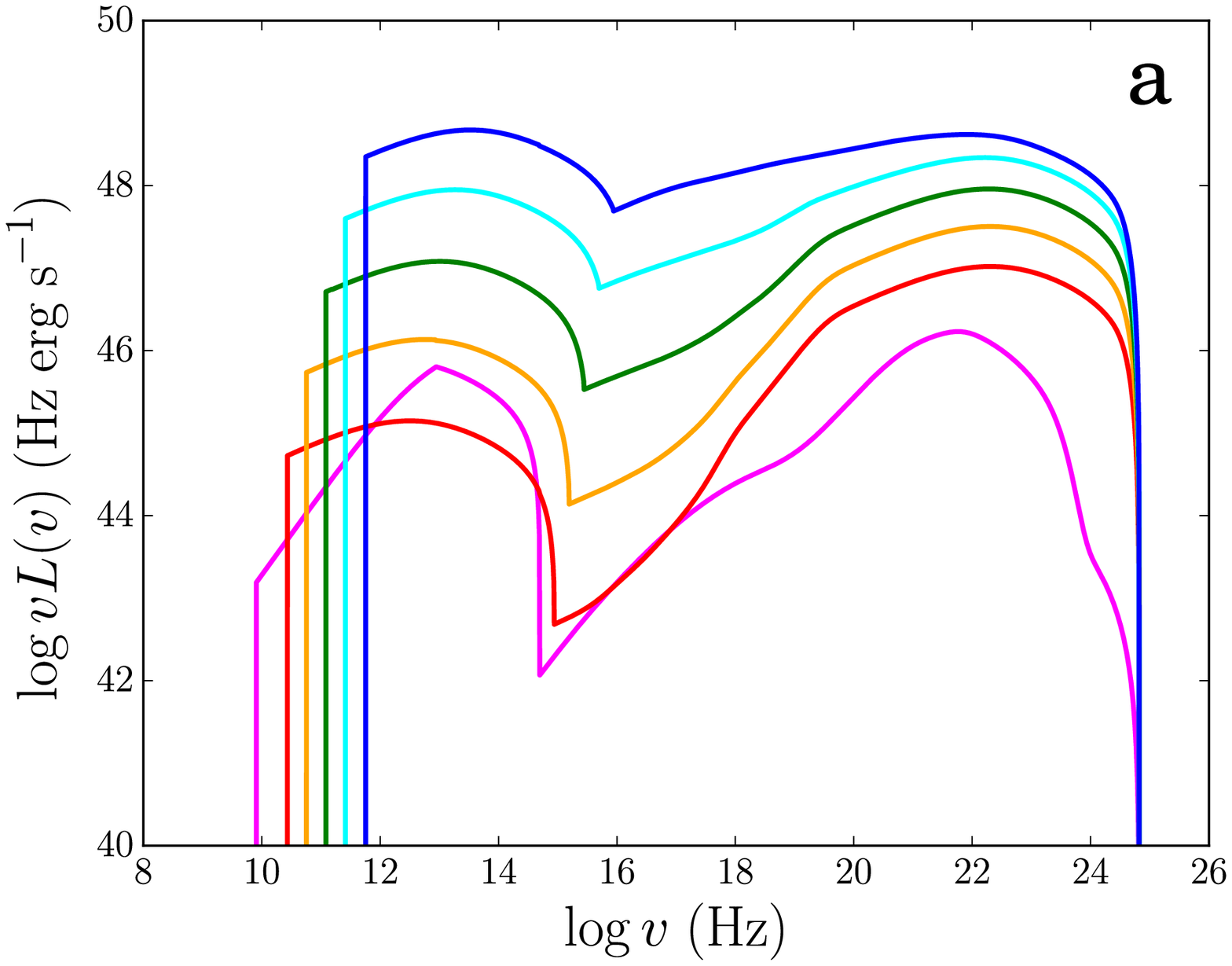} \\
\epsfxsize=8cm \epsfbox{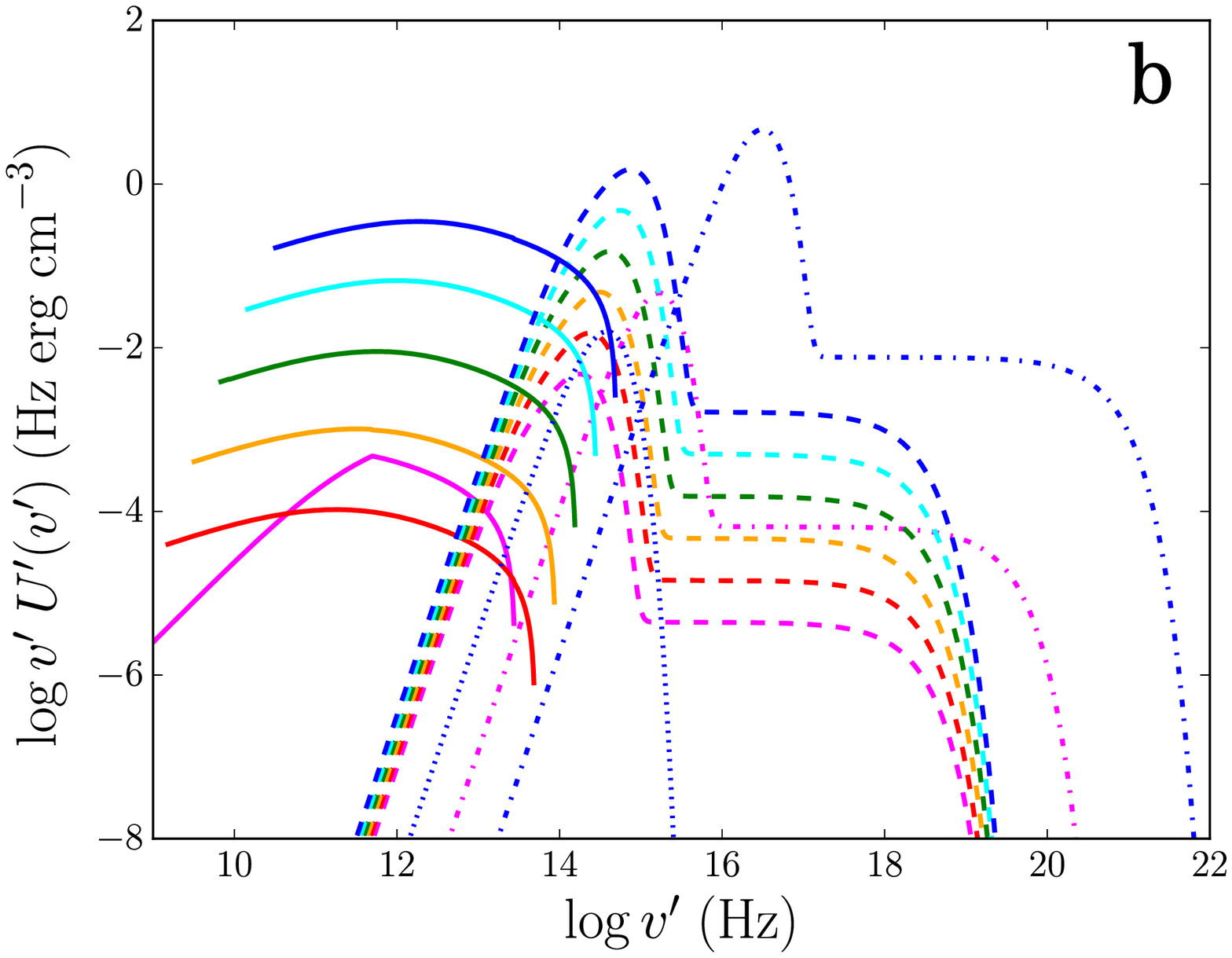} \\
\epsfxsize=8cm \epsfbox{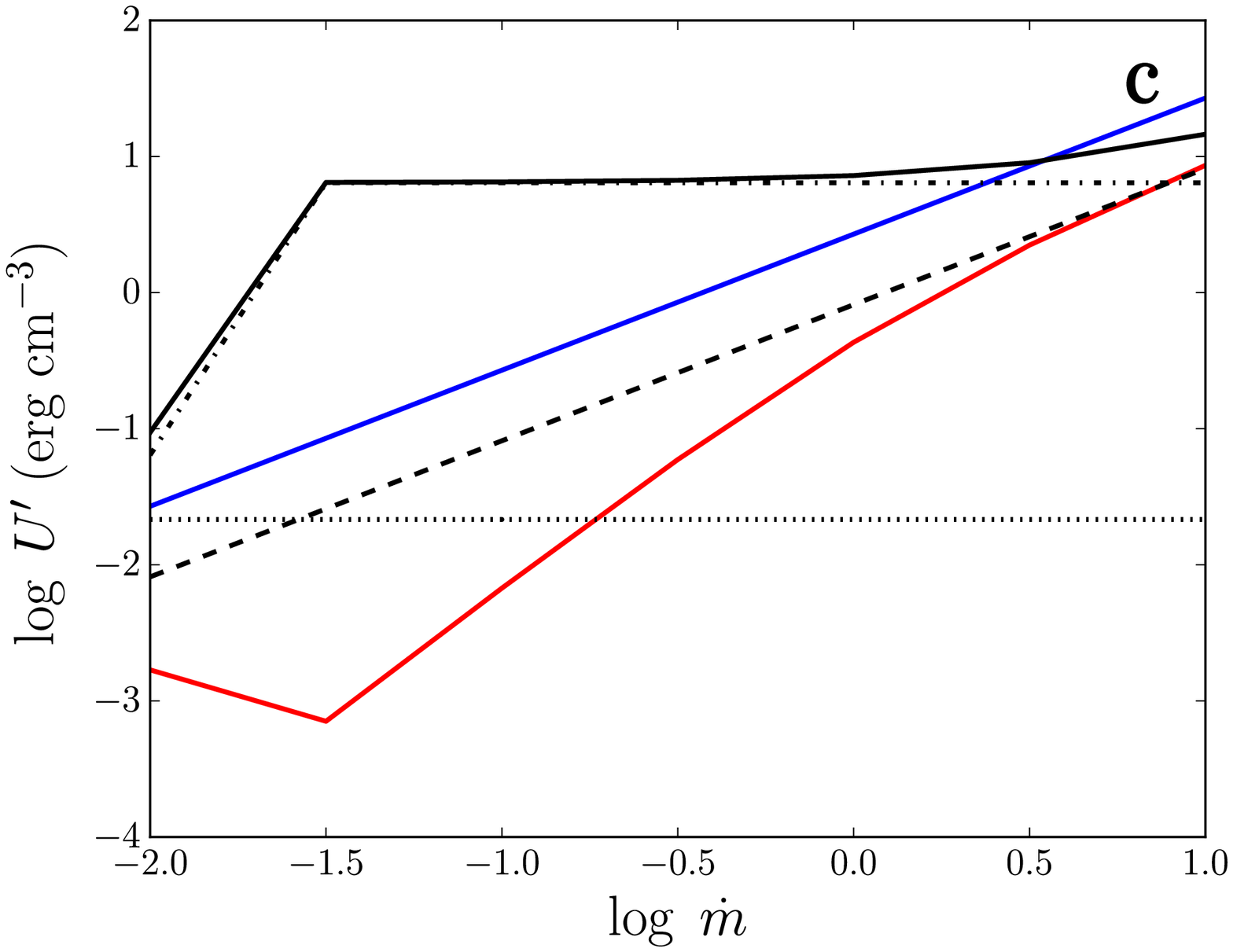} \\
\end{tabular}
\caption{Spectral changes with accretion rate using standard jet scalings ($R_{diss}\propto M$, $P^\prime_{rel}\propto \dot{m}M$, $B\propto (\dot{m}/M)^{1/2}$). a). FSRQ model SEDs for fixed BH mass and increasing accretion rate ($\log\dot{m}=-2$ (magenta), $-1.5$ (red), $-1$ (orange), $-0.5$ (green), $0$ (cyan) and $0.5$ (blue), $M_{BH}=10^9 M_\odot$). b). Corresponding seed photon energy density spectra as seen in the jet frame. Solid lines show synchrotron seed photons, dashed lines show accretion disc plus coronal seed photons, dot-dashed lines show seed photons from the BLR plus coronal flux reflected by the BLR, and dotted lines show seed photons from the torus. The seed photon energy density from the torus is the same for all accretion rates. The seed photon energy density from the BLR is the same for all accretion rates except $\log\dot{m}=-2$, where $Z_{diss}>R_{BLR}$. c). Seed photon energy densities in jet frame as a function of accretion rate. Blue line shows $U^\prime_B$, red line shows $U^\prime_{sync}$, black lines show energy densities of external seed photons: dashed line shows $U^\prime_d+U^\prime_X$, dot-dashed line shows $U^\prime_{BLR}+U^\prime_{XBLR}$, dotted line shows $U^\prime_{IR}$, and solid line shows total $U^\prime_{ex}$.}
\label{fig2}
\end{figure}

Fig.\ref{fig2}a shows a sequence of FSRQ spectra with increasing accretion rate. We fix $M=10^9$ and increase the accretion rate from $\log\dot{m}=-2$ to $0.5$, scaling $R_{diss}$, $B$ and $P^\prime_{rel}$ as described in \S \ref{sec:scalings}. 

\begin{figure*} 
\centering
\begin{tabular}{l|r}
\leavevmode  
\epsfxsize=8cm \epsfbox{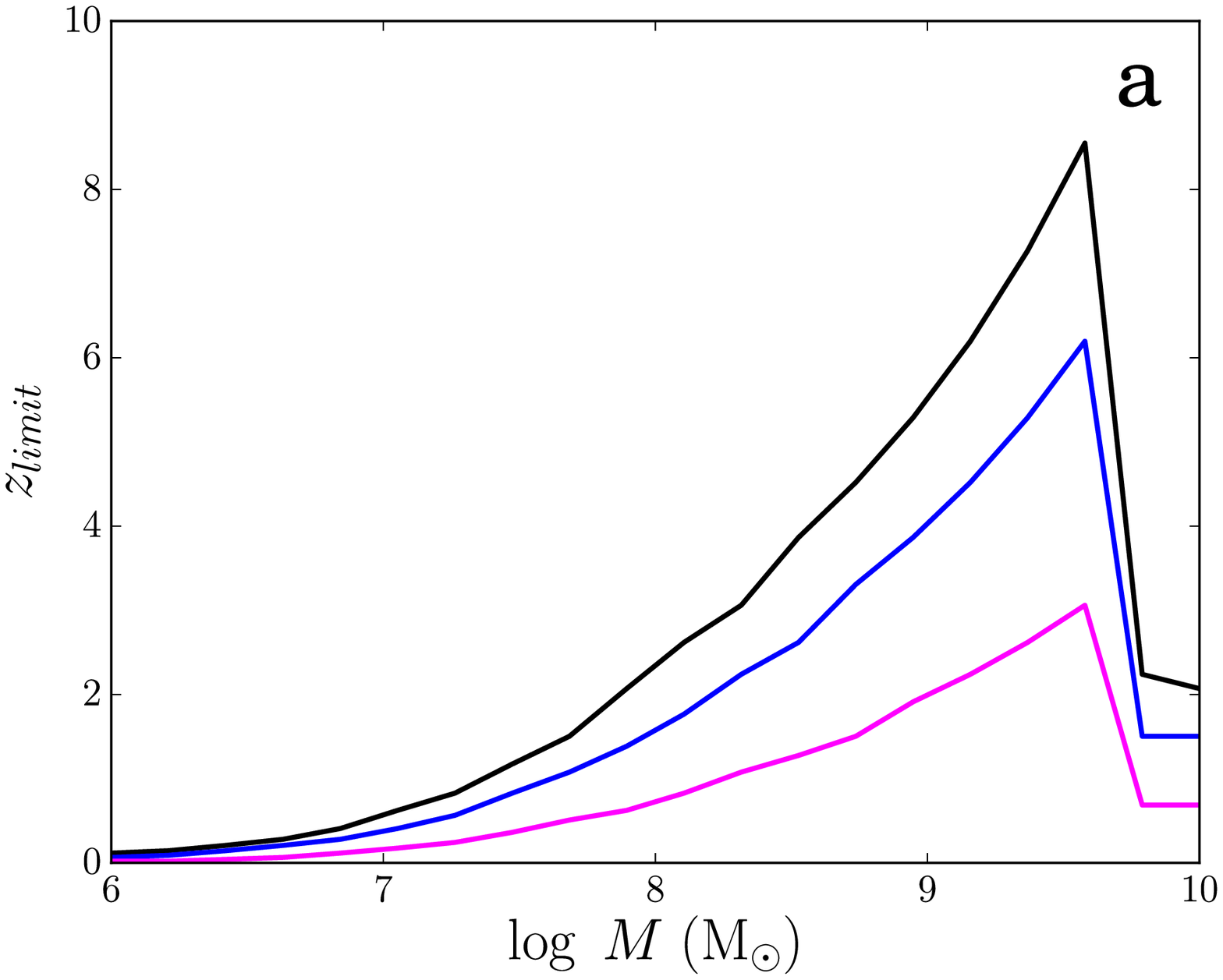} &
\epsfxsize=8cm \epsfbox{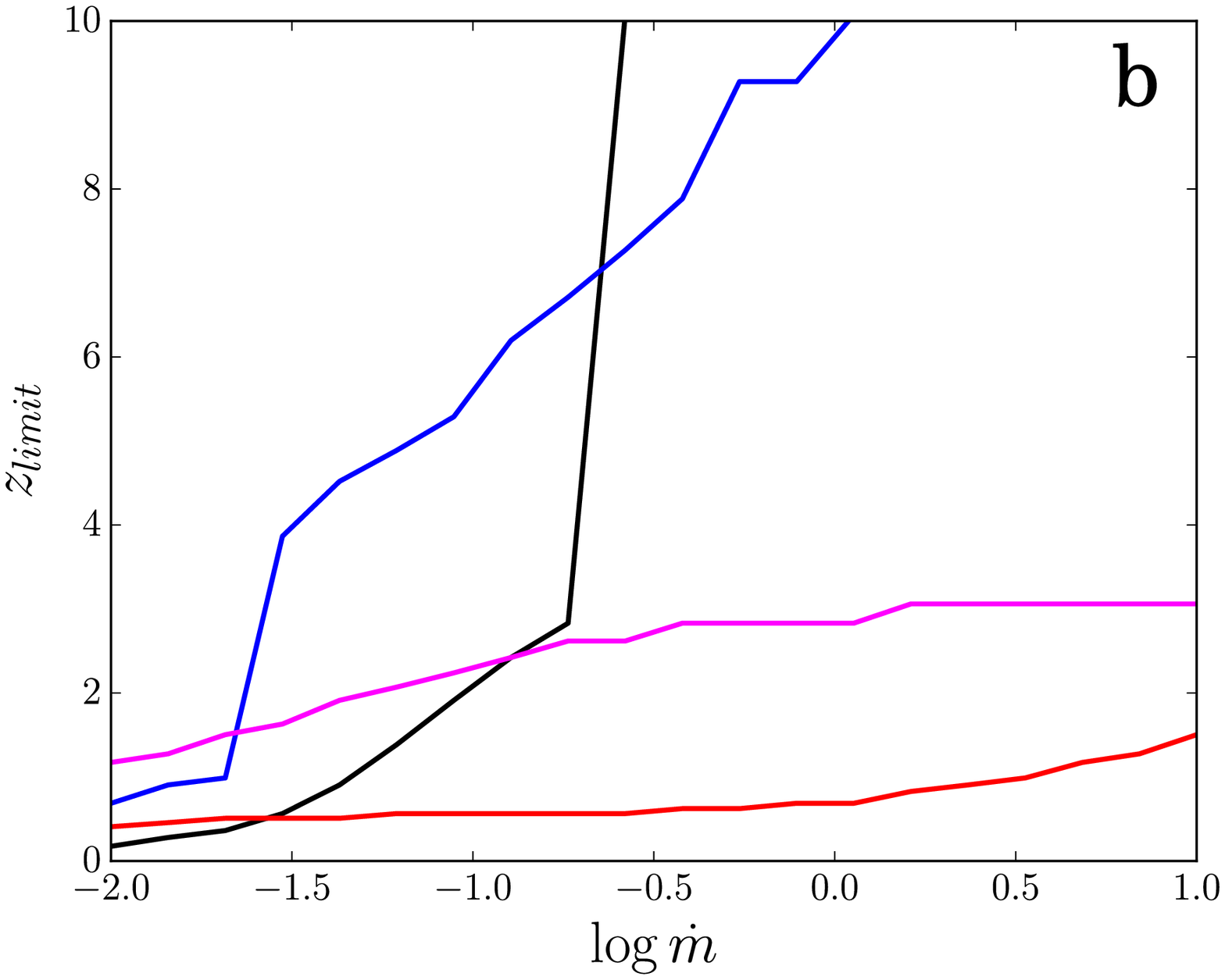} \\
\end{tabular}
\caption{a). Redshift limits for Fermi visible FSRQs as a function of black hole mass, for increasing viewing angle ($\theta=0$ (black), $1/\Gamma$ (blue) and $1/2\Gamma$ (magenta), where $\Gamma=13$) and $\dot{m}=0.1$. b). Redshift limits for Fermi visible FSRQs as a function of accretion rate, for $M_{BH}=10^7$ (red), $10^8$ (magenta), $10^9$ (blue) and $10^{10}M_{\odot}$ (black) and $\theta=0$.}
\label{fig3}
\end{figure*}

As accretion rate increases, the synchrotron self-absorption frequency increases from $\sim10^{10}$ (magenta) to $10^{12}$Hz (blue). This is because the size of the emission region stays constant ($R_{diss}$ does not depend on $\dot{m}$) while the magnetic field is increasing ($B\propto\dot{m}^{1/2}$). 

As accretion rate increases, the total luminosity also increases, since $P^\prime_{rel}\propto\dot{m}$ and $B\propto\dot{m}^{1/2}$. However, the synchrotron emission increases faster than the Compton emission, so that the two peaks show comparable luminosity for the highest accretion rate spectrum ($\log\dot{m}=0.5$; blue), while the Compton peak is $\sim2$ orders of magnitude brighter than the synchrotron peak at $\log\dot{m}=-1.5$ (red). 

This is because the increase in synchrotron emission comes from both the increase in $P^\prime_{rel}$ and the increase in its seed photons from the magnetic field. In contrast, the main source of seed photons for the Compton hump is the BLR and the energy density of BLR seed photons remains constant while $Z_{diss}<R_{BLR}$. Hence most of the increase in $L_{comp}$ is due to $P^\prime_{rel}$. Only for the highest accretion rates (blue and cyan spectra), do the other sources of seed photons ($U^\prime_{sync}$ and $U^\prime_{acc}$, solid and dashed lines in Fig.\ref{fig2}b) become comparable with $U^\prime_{BLR}$. These are much lower energy photons than the blue shifted BLR emission (Fig.\ref{fig2}b). Consequently the Compton hump at the highest accretion rates is much broader, as well as being more similar in luminosity to the synchrotron peak. Again this gives a spectral shape much more typical of low accretion rate BL Lacs, except now it is the result of an extremely high $\dot{m}$ causing $U^\prime_{sync}$ to dominate over $U^\prime_{ex}$, rather than (in the case of BL Lacs) an extremely low $\dot{m}$ where $U^\prime_{ex}$ is absent. 

The red spectrum in Fig.\ref{fig2}a ($\log\dot{m}=-1.5$) shows the greatest luminosity difference between synchrotron and Compton peaks. The difference lessens again for the lowest accretion rate spectrum (magenta, $\log\dot{m}=-2$). This is because $R_{BLR}\propto L_d^{1/2}\propto\dot{m}^{1/2}$. For a $10^9M_\odot$ BH at $\dot{m}=10^{-2}$, the BLR radius has shrunk so much that it is now less than the distance to the jet emission region. Once the BLR is behind the jet emission region, its seed photons are de-boosted and $U^\prime_{BLR}$ drops significantly (compare magenta and blue dot-dashed lines in Fig.\ref{fig2}b). The amount of cooling drops, shown by the appearance of a cooing break at $10^{13}$Hz in the magenta spectrum. Synchrotron and accretion flow seed photons become more important, broadening the Compton hump again. But even these cannot help for long; $\dot{m}=10^{-2}$ is the rate at which accretion flows make the transition from radiatively efficient to inefficient. Below $\dot{m}=10^{-2}$ UV bright accretion discs can no longer be sustained and give way to ADAF like flows. This severely reduces the available accretion flow seed photons and effectively switches off the BLR, since there are no UV photons to illuminate it. This final magenta spectrum represents the transition from dimming FSRQ to a low accretion rate SSC BL Lac. 

Fig.\ref{fig2}c shows the total energy densities of seed photons in the jet frame as a function of accretion rate. For $-1.5<\log\dot{m}<-0.5$, $U^\prime_{BLR}$ dominates, suppressing $U^\prime_{sync}$ and giving the luminous Compton hump and much smaller synchrotron peak typical of FSRQs. Only at the extremes of accretion rate does $U^\prime_{sync}$ dominate. At super Eddington accretion rates, $U^\prime_{sync}$ and $U^\prime_{acc}$ start to overtake $U^\prime_{BLR}$, producing a pseudo-BL Lac type spectrum but with extremely high luminosity. At the lowest accretion rates, $U^\prime_{sync}$ recovers when $R_{BLR}$ has shrunk below $Z_{diss}$, reducing the external seed photons and beginning the object's transition to a low accretion rate SSC BL Lac.

\section{FSRQ Visibility}

\begin{figure*} 
\centering
\begin{tabular}{l|c|r}
\leavevmode  
\epsfxsize=5cm \epsfig{file=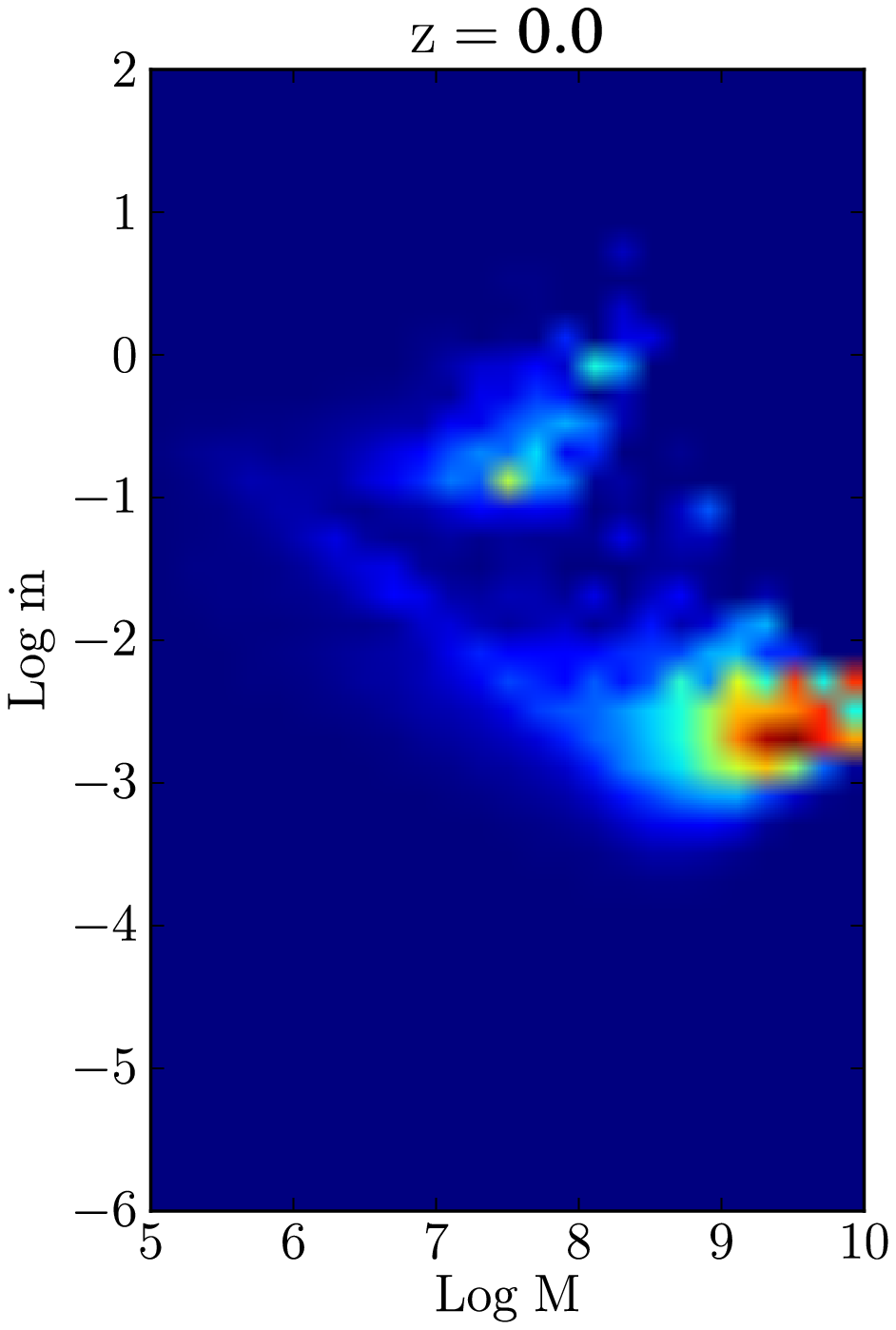,clip=,width=6cm} &
\epsfxsize=5cm \epsfig{file=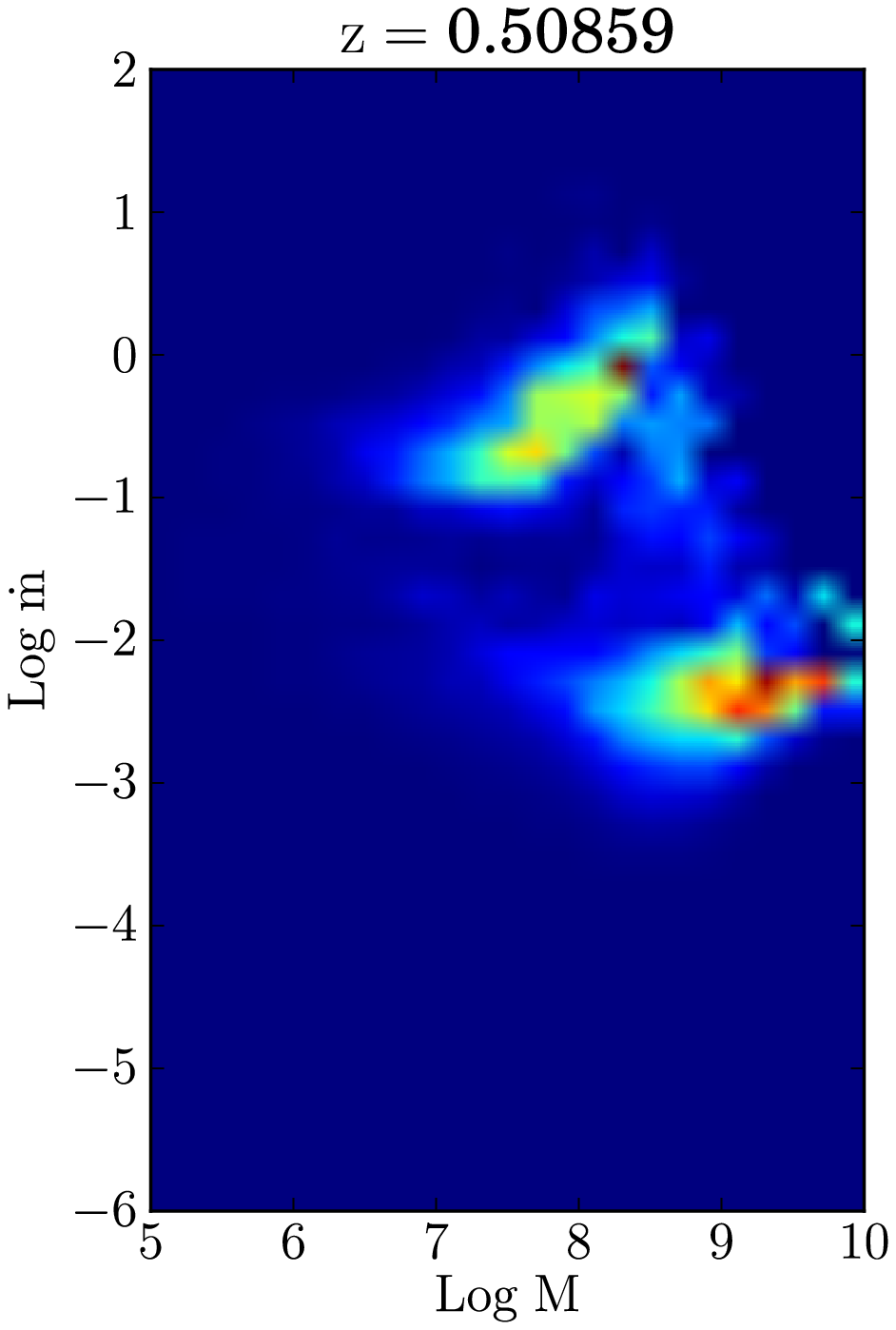,clip=,width=6cm} &
\epsfxsize=5cm \epsfig{file=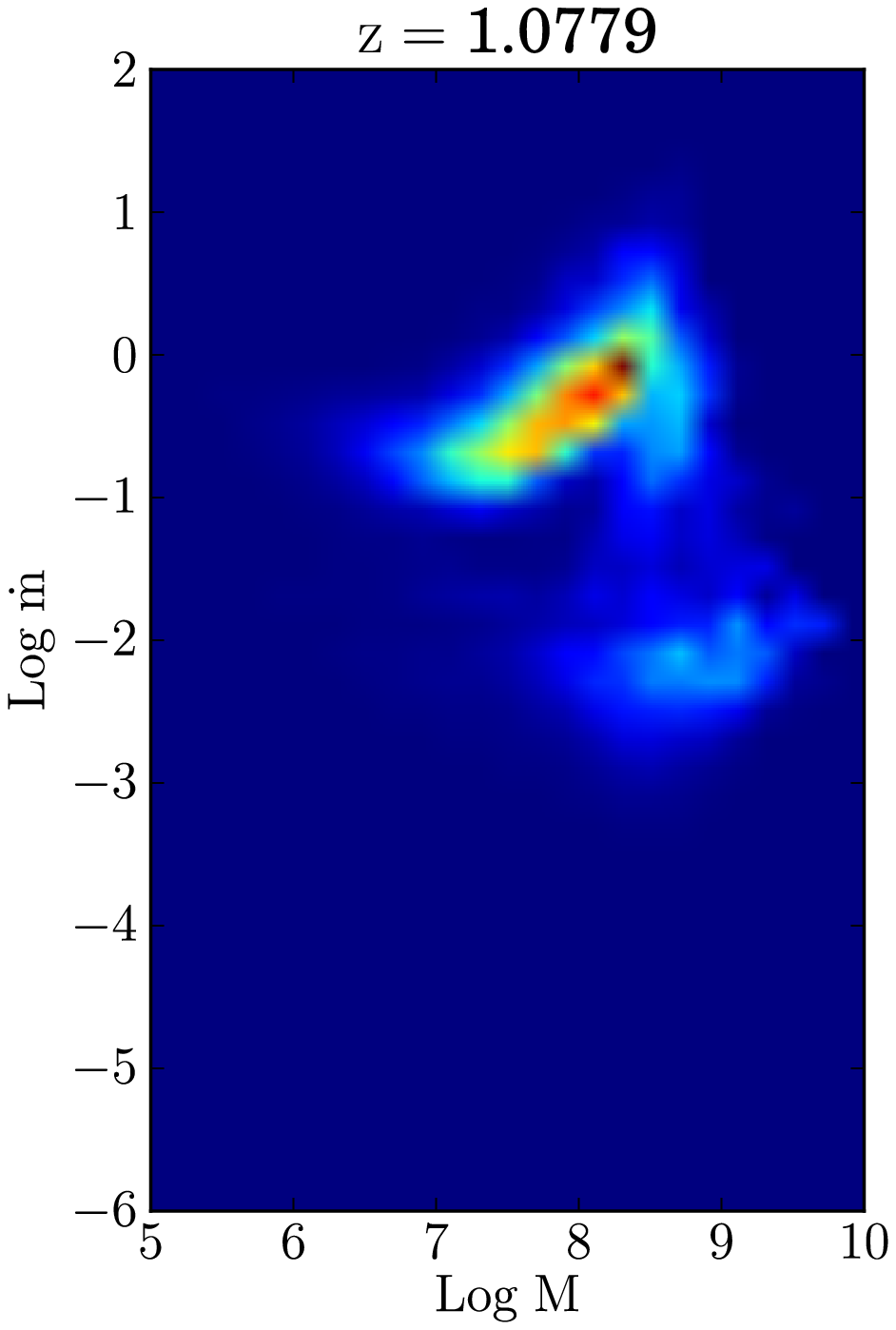,clip=,width=6cm} \\
\epsfxsize=5cm \epsfig{file=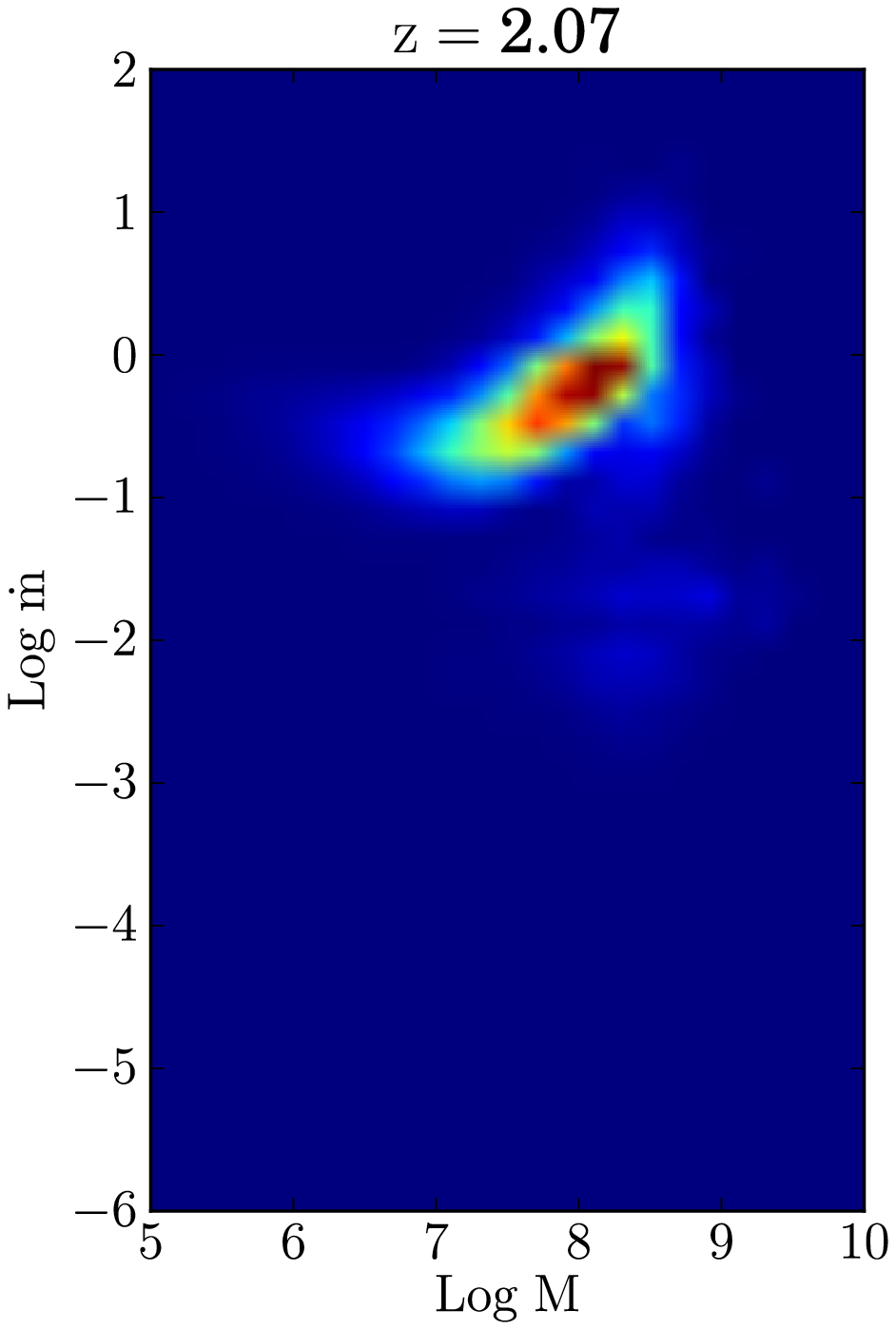,clip=,width=6cm} &
\epsfxsize=5cm \epsfig{file=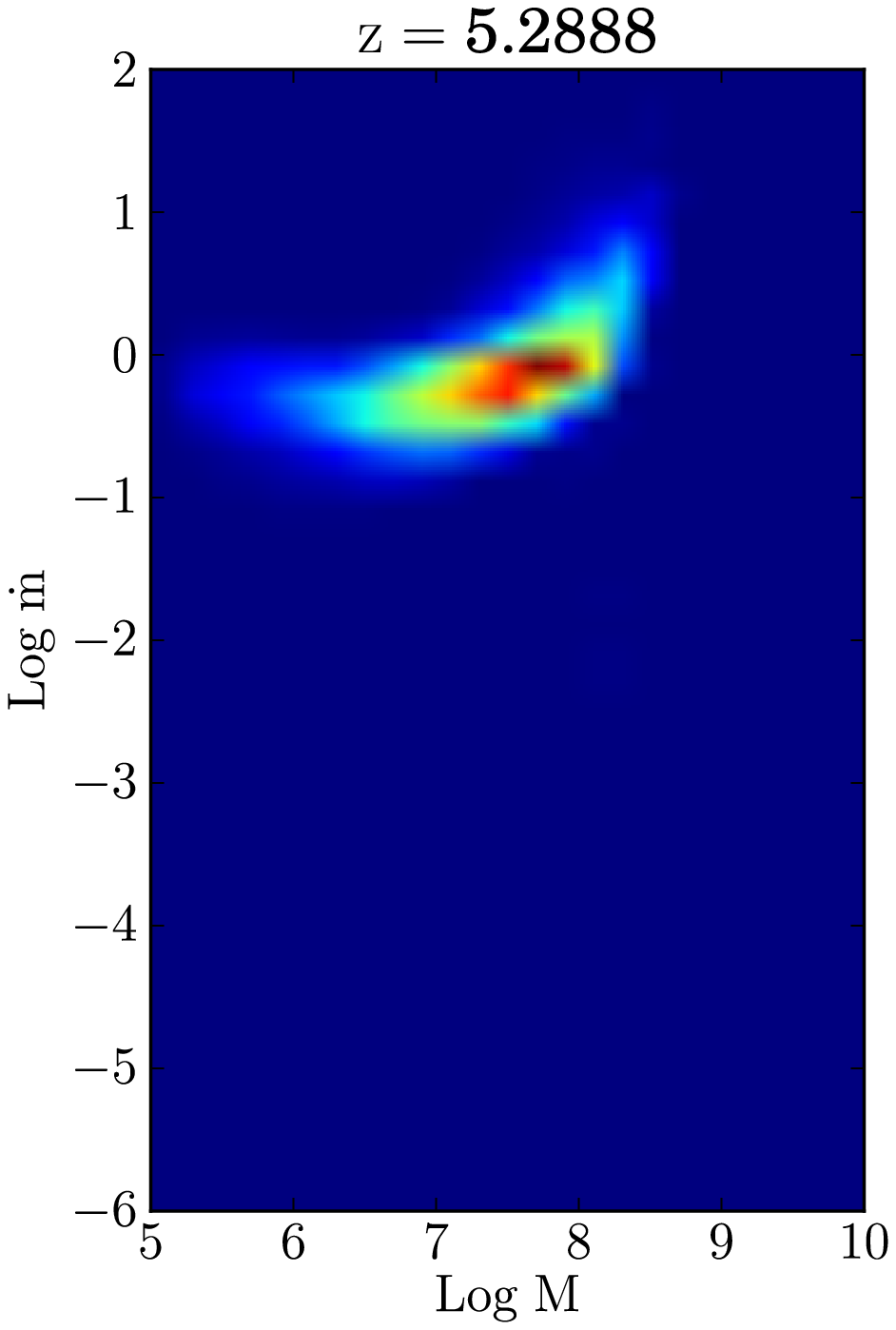,clip=,width=6cm} &
\epsfxsize=5cm \epsfig{file=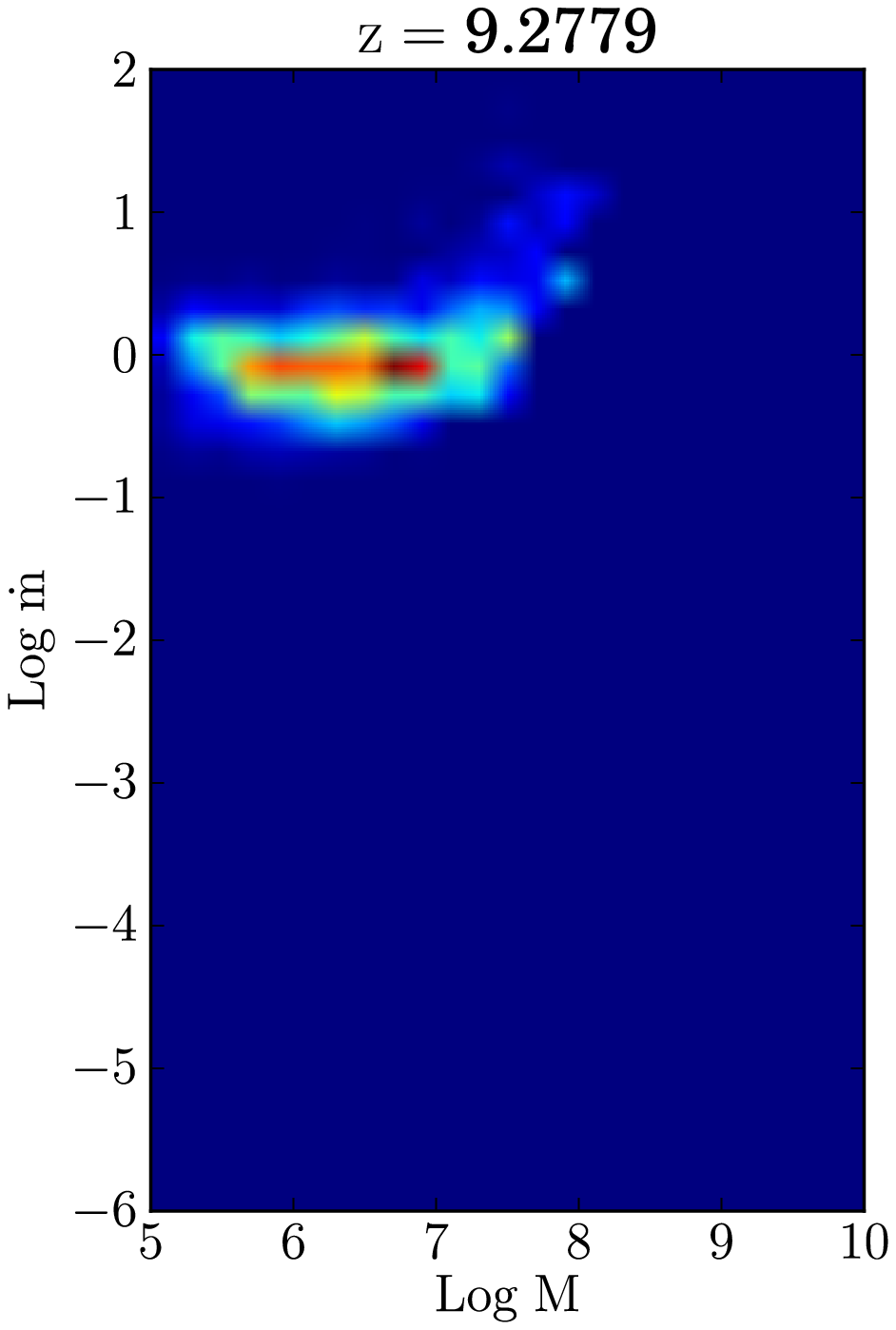,clip=,width=6cm} \\
\end{tabular}
\caption{Predicted mass and accretion rate distribution of accreting black holes at increasing redshift from the Millennium simulation. Colours trace luminosity density, where we define luminosity density as the number density multiplied by the luminosity ($L$) at that mass and mass accretion rate, where $L=\eta \dot{M} c^2$ for the thin disc regime ($10^{-2}<\dot{m}<1$), joining smoothly onto a radiatively inefficient regime at lower $\dot{m}$ where $L\propto \dot{m}^2$ (Narayan \& Yi 1995) and onto a super-Eddington flow at higher $\dot{m}$ where $L\propto \ln(1+\dot{m})$ (Shakura \& Sunyaev 1973). The luminosity density in each $(z,M,\dot{m})$ bin therefore depends on the mass, accretion rate, spin (which sets $\eta$), the inferred accretion regime and the number of black holes in that bin. Red shows the mass and accretion rates at which the maximum accretion luminosity is emitted at each redshift.}
\label{fig3.5}
\end{figure*}

Having shown how the spectrum of a FSRQ might change with mass and accretion rate, we now investigate the redshift limits at which FSRQs of different masses and accretion rates should be visible to Fermi. We define a flux limit of $F_{1GeV-100GeV}>5\times10^{-10} photons\,cm^{-2} s^{-1}$ from the Fermi 1 year catalogue (Abdo et al. 2010). If a FSRQ of a given mass and accretion rate has $F>F_{limit}$ in the Fermi band we assume it will be detected. 

Fig.\ref{fig3}a shows the redshift limits for Fermi visible FSRQs as a function of BH mass. We fix $\dot{m}=0.1$ and show three different inclination angles: $\theta=0$ (black), $1/\Gamma$ (blue) and $1/2\Gamma$ (magenta). Clearly more closely aligned FSRQs are seen out to higher redshifts. The limiting redshift increases with mass, since $L_{comp}$ increases with mass (see Fig.\ref{fig1}a), until $\sim10^{9.6}M_\odot$. A highly aligned FSRQ with a $10^{9.5}M_\odot$ BH can be detected out beyond $z=6$. However, above $10^{9.5}M_\odot$, the redshift limits drop sharply to $z\leqslant 2$ for a $10^{10}M_\odot$ BH. The reason for this can be seen in Fig.\ref{fig1}a. For the most massive $10^{10}M_\odot$ BHs, $Z_{diss}>R_{BLR}$, because $Z_{diss}$ grows $\propto M$ while $R_{BLR}\propto M^{1/2}$. BLR photons are still the dominant source of seed photons, however they are now behind the emission region. Consequently they are deboosted, so that the peak energy of BLR seed photons is lower. This shifts the peak of the Comptonised emission to lower energies and the flux in the Fermi band ($1-100$GeV, corresponding to $23.38<\log\nu<25.38$) drops significantly. The luminosity of a $10^{10}M_\odot$ FSRQ at $10^{24}$Hz is almost 2 orders of magnitude less than a $10^9M_\odot$ BH at the same accretion rate (compare black and blue lines, Fig.\ref{fig1}a). Redshifting the spectrum only exacerbates the shift of the Compton peak to lower energies and further reduces the Fermi flux. Consequently, the redshift limits of $10^{10}M_\odot$ FSRQs are nearer those of $10^{7-8}M_\odot$ BHs. 

Fig.\ref{fig3}b shows the redshift limits for FSRQs as a function of accretion rate for four different BH masses ($M_{BH}=10^7$ (red), $10^8$ (magenta), $10^9$ (blue) and $10^{10}M_{\odot}$ (black) and $\theta=0$). $z_{limit}$ increases with $\dot{m}$, however the rate of increase differs with mass. 

The redshift limits for $10^{7-8}M_\odot$ FSRQs increase very slowly with accretion rate (magenta and red lines). The redshift limit for a $10^7M_\odot$ FSRQ is $\sim0.5$ at $\log\dot{m}=-2$ and $\sim0.75$ at $\log\dot{m}=1$. $10^8M_\odot$ FSRQs show a similarly small factor $\sim3$ increase over the same range in accretion rate. This is because the dominant cooling is through SSC for low mass FSRQs, due to the small emission region size and high magnetic field. As a low mass FSRQ ($10^{7-8}M_\odot$) increases its accretion rate from $\log\dot{m}=-2$ to $1$, its Compton spectrum changes from being high peaked (at $\sim10^{24}$Hz) to low peaked ($\sim10^{21}$Hz), analogous to the change in BL Lac spectra from high peaked to low peaked. The reason is the same: increasing $\dot{m}$ increases the cooling, shifting all the peak energies to lower frequency, because low mass FSRQs are similarly dominated by SSC cooling (plus low energy accretion disc seed photons), which always dominates over BLR IC. Even though the total luminosity is increasing, the shift of the peak emission to lower energies means the Fermi band flux increases more slowly and hence $z_{limit}$ shows a very gradual increase. 

In contrast, $10^9M_\odot$ FSRQs show a much faster increase in $z_{limit}$ with $\dot{m}$ (blue line, Fig.\ref{fig3}b). This is because they are almost always dominated by BLR Compton scattering. The spectral energy density of BLR seed photons peaks at higher energy (see Fig.\ref{fig1}b, blue dot-dashed line), than the synchrotron and disc seed photons which dominate in lower mass systems, hence the IC peak is at higher energy ($10^{23}$ compared to $10^{20}$Hz, compare blue and magenta lines, Fig.\ref{fig1}a), so more of the luminosity increase can be seen in the Fermi band. Only at the very lowest accretion rates ($\log\dot{m}\sim-2$) does the Fermi visibility of a $10^9M_\odot$ FSRQ dip below that of a $10^8M_\odot$ object. This is because the Compton cooling is slightly more efficient in the larger mass object, shifting its Compton peak to slightly lower energy and hence giving it a lower Fermi band flux. 

The $10^{10}M_\odot$ FSRQ (black line, Fig.\ref{fig3}b) shows a similar effect, with $z_{limit}$ increasing slowly at first and then more rapidly for $\log\dot{m}>-1$. This is because $R_{BLR}\propto M^{1/2}\dot{m}^{1/2}$ while $Z_{diss}\propto M$, hence for larger mass a higher $\dot{m}$ is needed for $R_{BLR}>Z_{diss}$, i.e. $R_{BLR}<Z_{diss}$ up to higher $\dot{m}$. While $R_{BLR}<Z_{diss}$, the IC hump is dominated by synchrotron, accretion flow and deboosted BLR seed photons, so its peak is at lower frequency and the Fermi band flux ($\sim10^{24}$Hz) is significantly reduced. Once $R_{BLR}>Z_{diss}$ (at $\log\dot{m}\sim-1$ for $M=10^{10}M_\odot$), Doppler boosted BLR seed photons dominate and $z_{limit}$ increases dramatically. 

\begin{figure} 
\centering
\begin{tabular}{l}
\leavevmode  
\epsfxsize=8cm \epsfbox{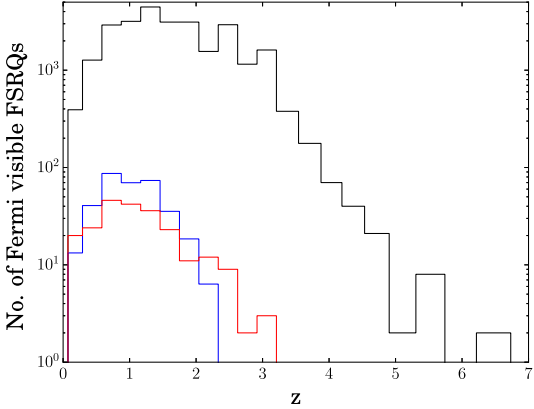} \\
\end{tabular}
\caption{Comparison of observed and predicted FSRQ redshift distributions. Red line shows observed redshift distribution of Fermi detected FSRQs found by Shaw et al. (2012). Black line shows predicted redshift distribution assuming all BHs with $\dot{m}>0.01$ produce a FSRQ type jet. Blue line shows predicted redshift distribution of Fermi visible FSRQs, assuming only BHs with $\dot{m}>0.01$ and spin $a>0.77$ produce a FSRQ type jet.}
\label{fig4}
\end{figure}

\section{Predicted FSRQ Population from Cosmological Simulations}

\begin{figure*} 
\centering
\begin{tabular}{l|r}
\leavevmode  
\epsfxsize=7cm \epsfbox{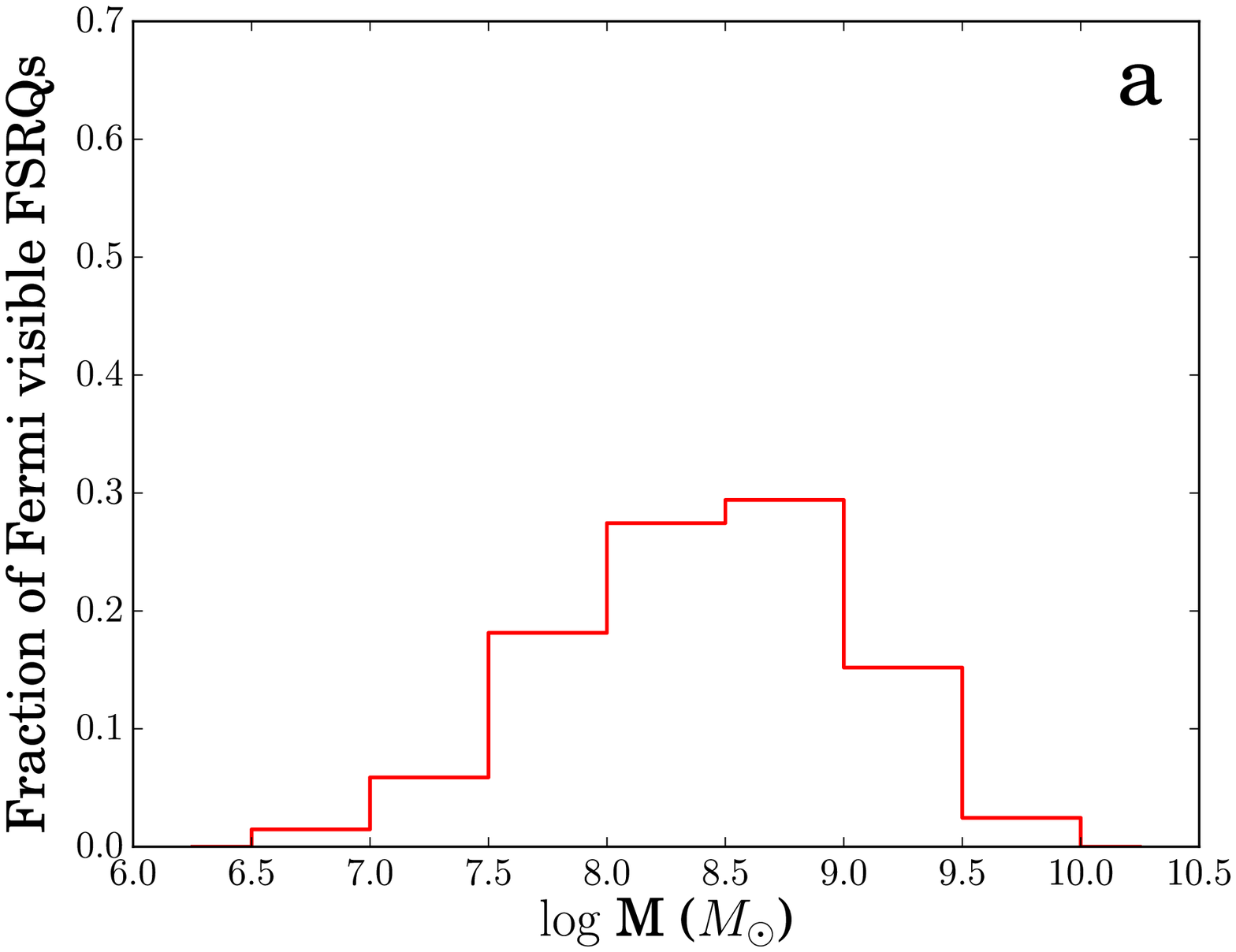} &
\epsfxsize=7cm \epsfbox{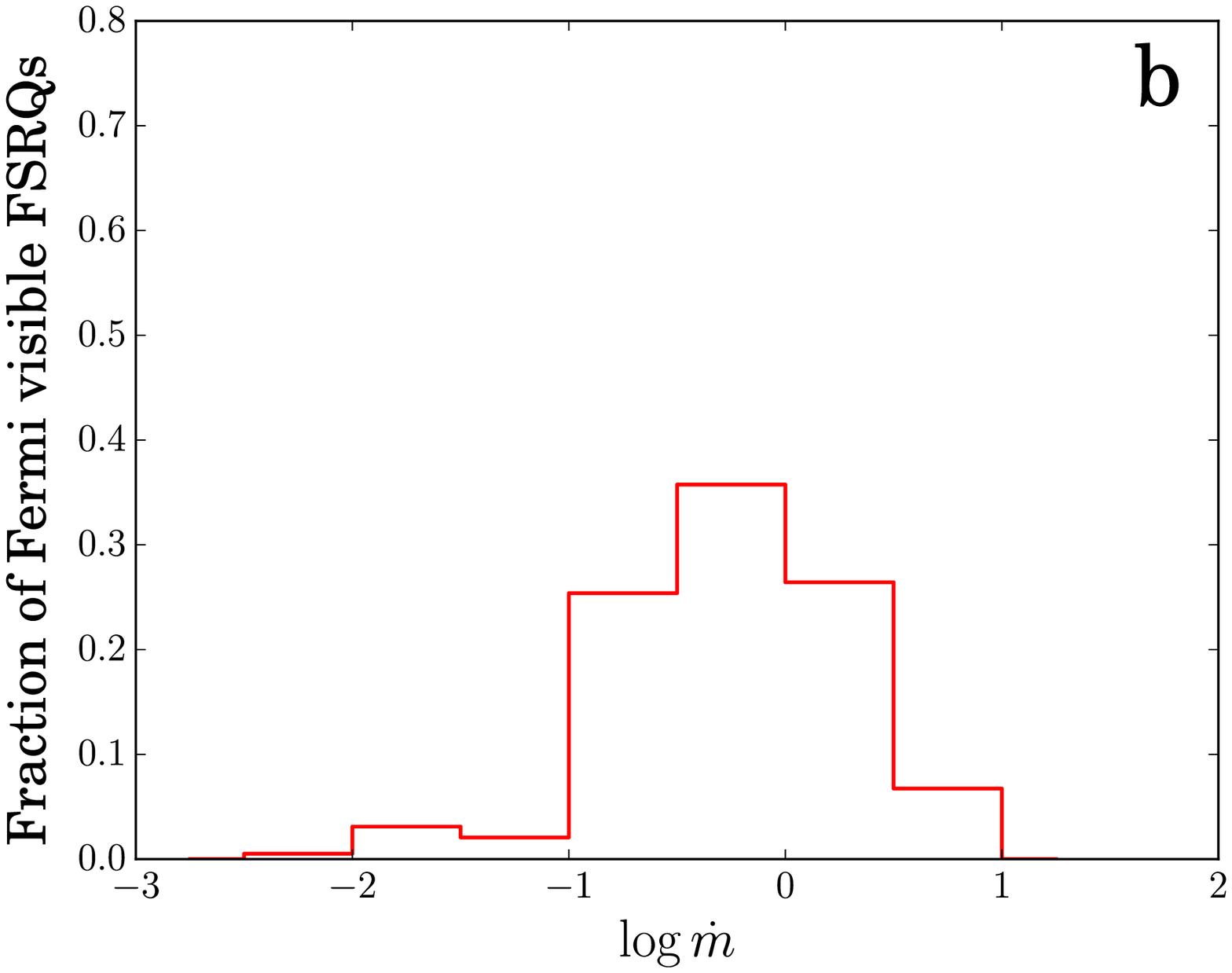} \\
\epsfxsize=7cm \epsfbox{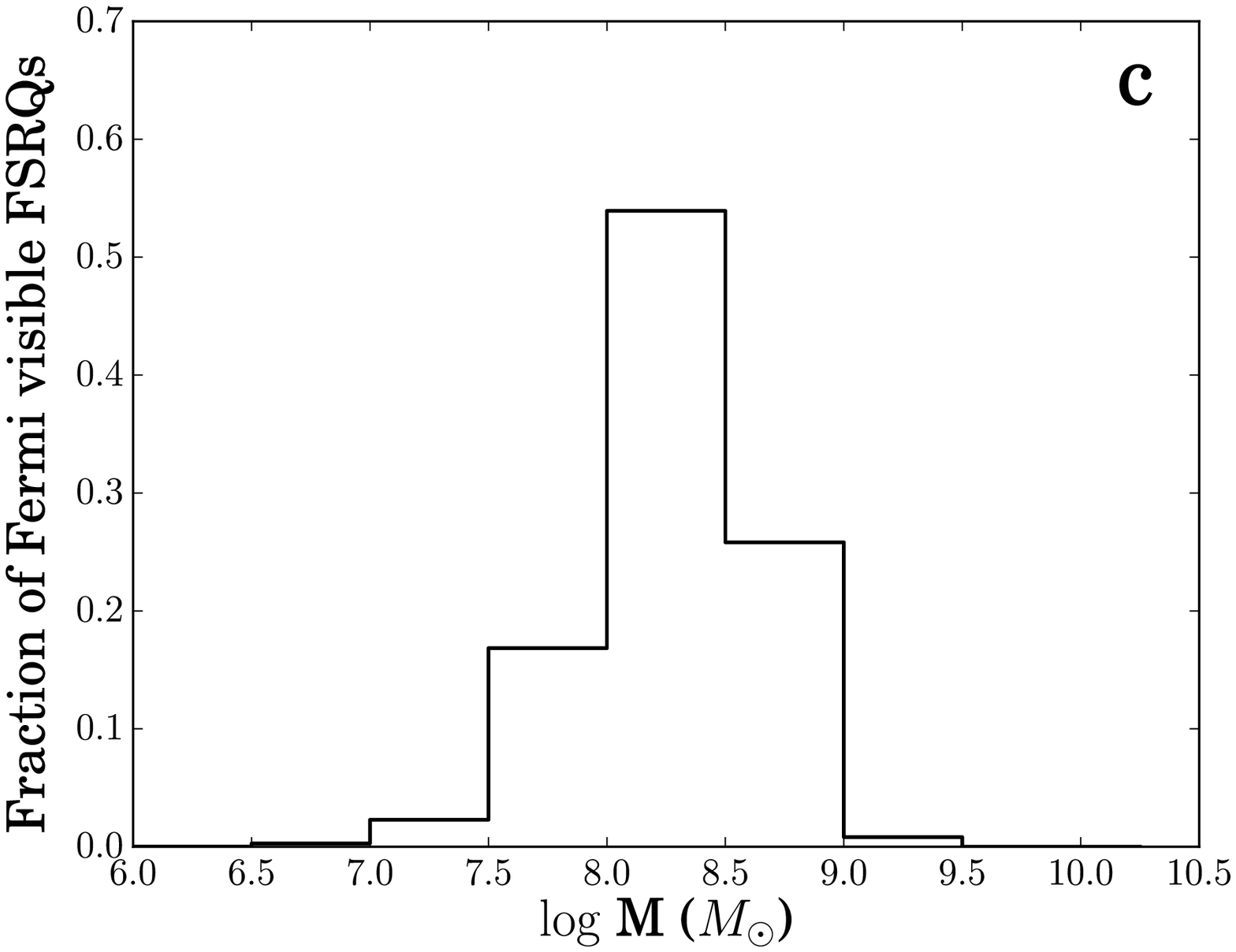} &
\epsfxsize=7cm \epsfbox{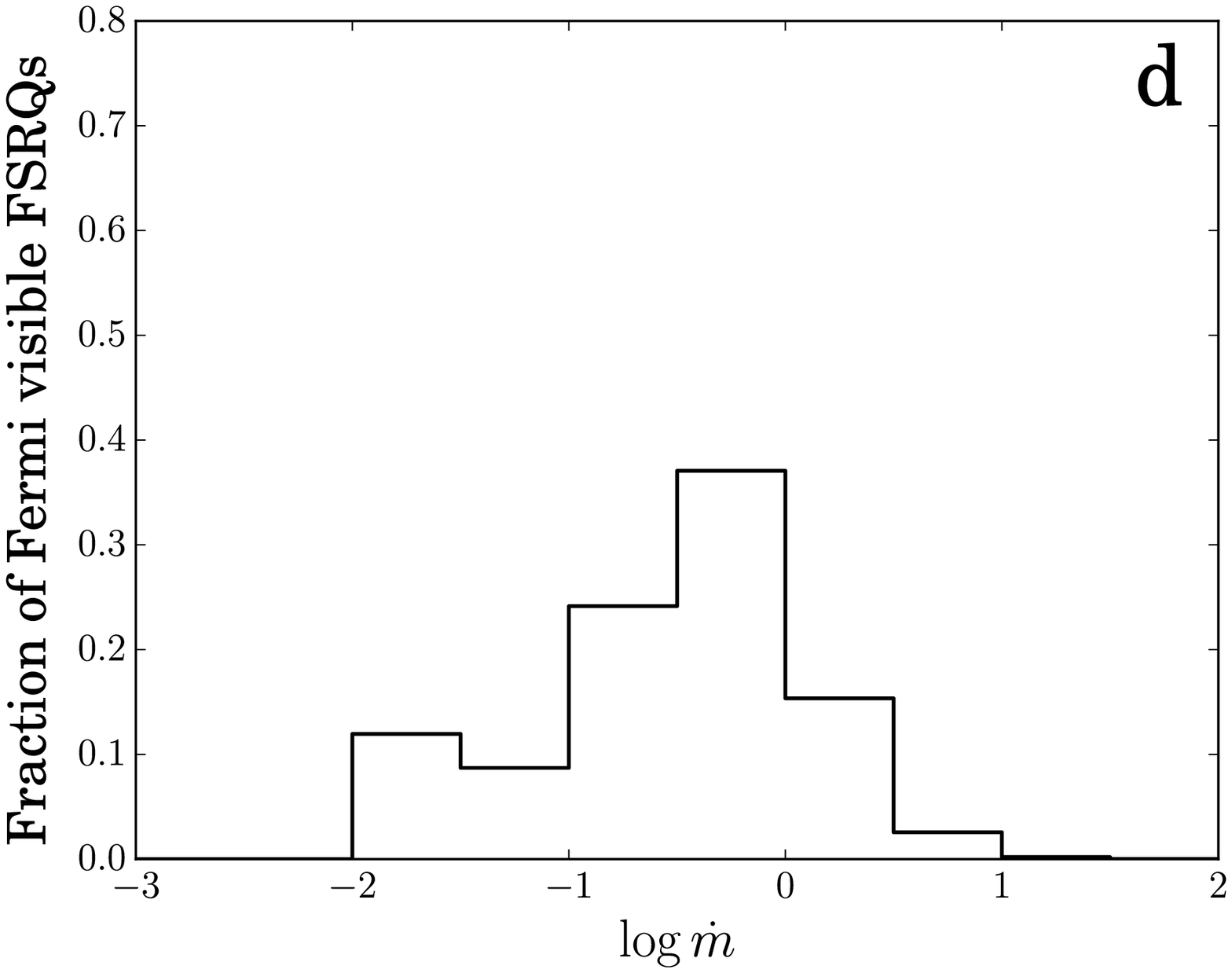} \\
\epsfxsize=7cm \epsfbox{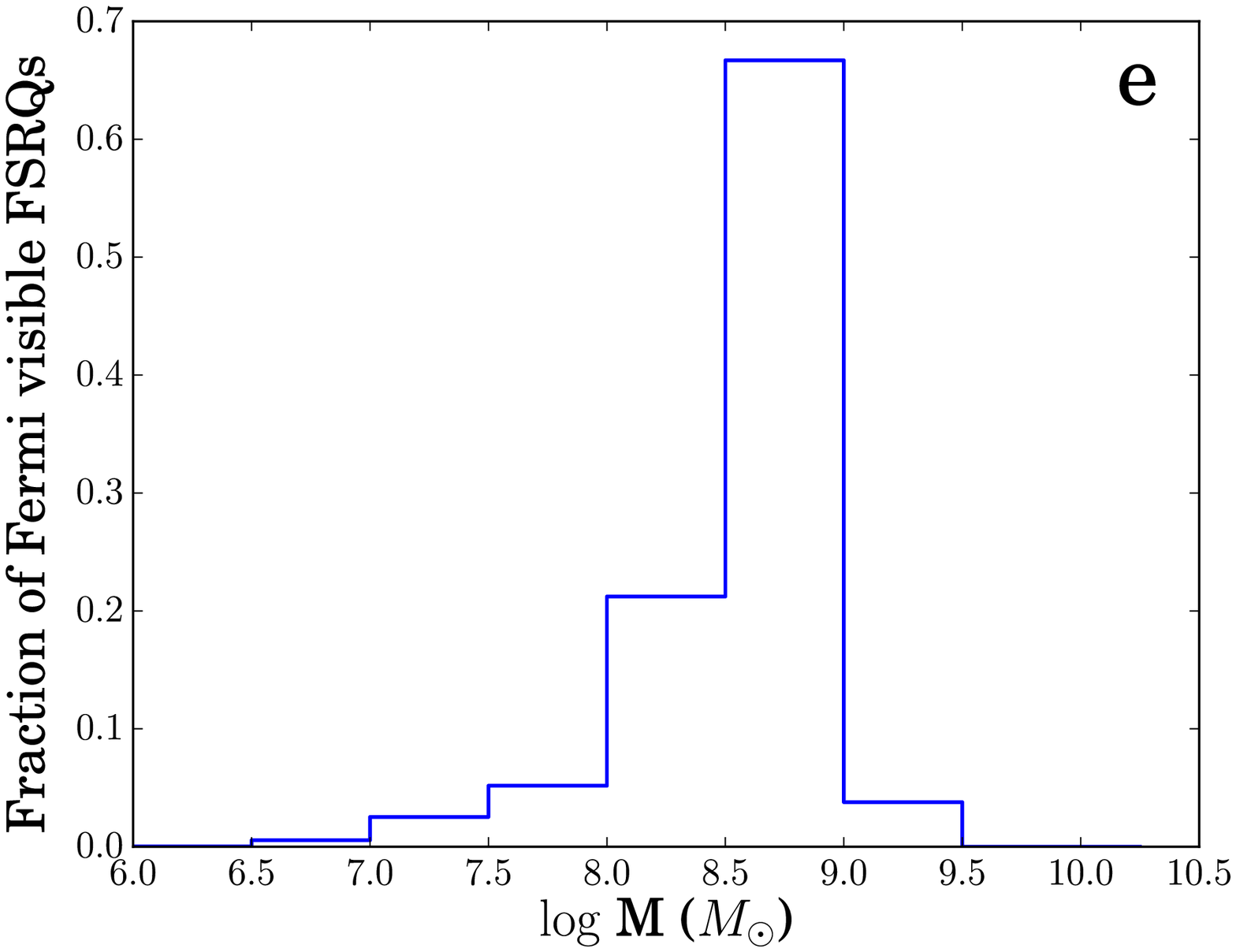} &
\epsfxsize=7cm \epsfbox{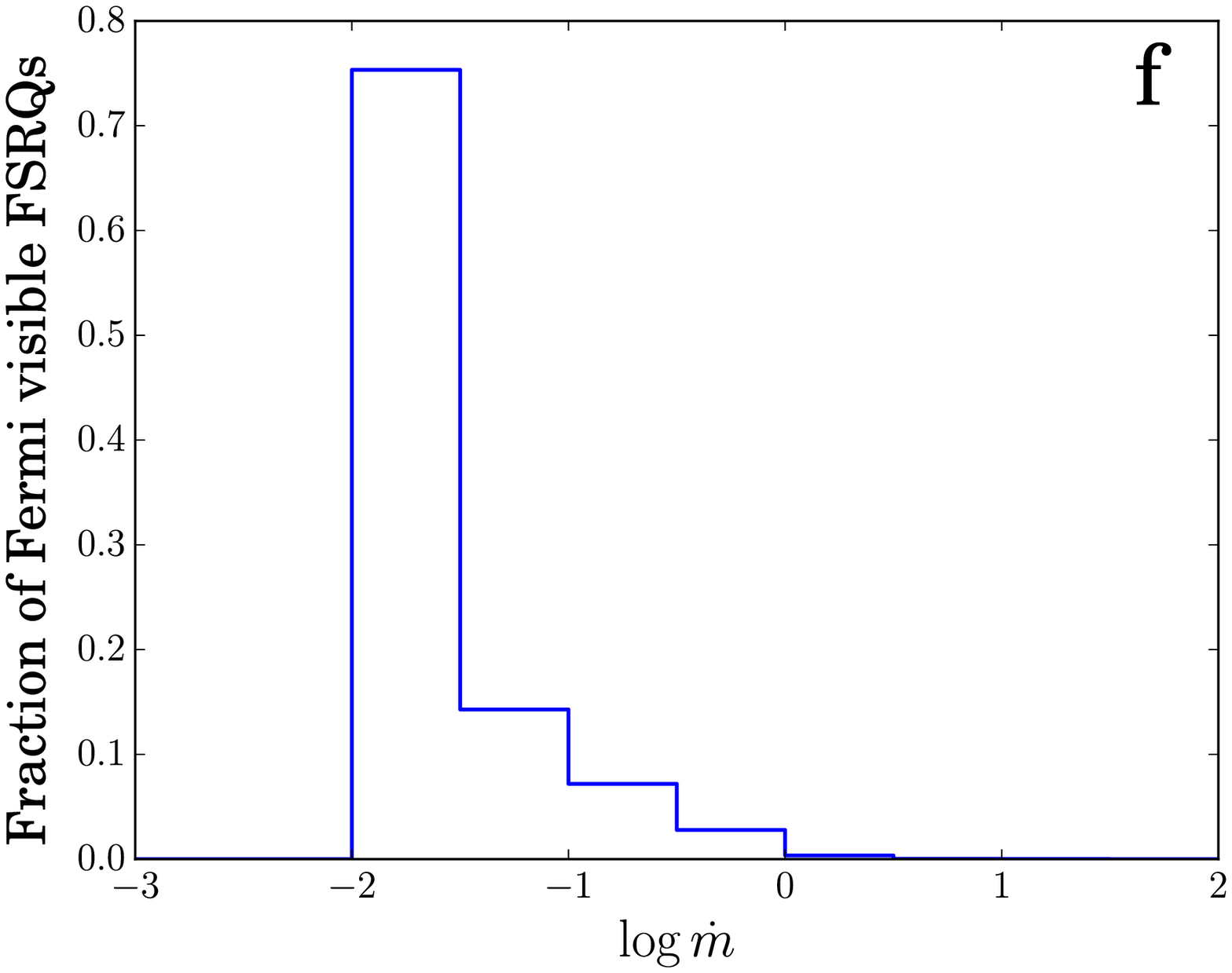} \\
\end{tabular}
\caption{a). and b). show, respectively, observed mass and accretion rate distributions of Fermi visible FSRQs from the data of Shaw et al. (2012). c). and d). show predicted mass and accretion rate distributions of Fermi visible FSRQs assuming all BHs with $\dot{m}>0.01$ produce a FSRQ type jet. e). and f). show predicted mass and accretion rate distributions of Fermi visible FSRQs assuming only BHs with $\dot{m}>0.01$ and $a\geq0.77$ produce a FSRQ type jet.}
\label{fig5}
\end{figure*}

Combining our scaled jet emission model with the results from cosmological simulations allows us to predict the population of FSRQs that should be detected by Fermi. As in Paper 1, we use the BH number densities from the Millennium Simulation (Springel et al. 2005; Fanidakis et al. 2011; 2012), which predict the number of SMBHs accreting at different redshifts together with their masses and accretion rates ($n(z,M,\dot{m})$). In Fig.\ref{fig3.5} we show the luminosity density (ie. number density multiplied by accretion luminosity) of BHs predicted by the simulation as a function of mass and accretion rate at different redshifts. The simulation has been found to agree well with the observed number densities of broad line and narrow line AGN in the local universe (Fanidakis et al. 2011; 2012).

We initially assume that all BHs accreting with $\dot{m}>10^{-2}$, i.e. in the radiatively efficient regime, produce a FSRQ type jet. We can then calculate the number of AGN hosting a FSRQ jet in each $(z,M,\dot{m})$ bin. If this number is less than 1 we use Poisson statistics to randomly determine whether a BH is present or not. Each BH in each $(z,M,\dot{m})$ bin is then assigned a random distance within this redshift bin and random $\theta_{obs}$, assuming $\cos\theta_{obs}$ is distributed uniformly. We then calculate the observed spectrum to determine whether or not the jet would be visible to Fermi. We choose the flux limit of the Fermi 1 year catalogue, in order to compare our simulation results with the observations presented in Shaw et al. (2012). 

Fig.\ref{fig4} shows the predicted redshift distribution of Fermi visible FSRQs (black line). The predicted distribution peaks between redshifts $1<z<2.5$. This corresponds to the peak in quasar activity at $z\sim2$. At later times ($z<1$), typical BH accretion rates drop below $10^{-2}$ due to systems running out of gas to accrete (Fig.\ref{fig3.5}). At low accretion rates, the accretion flow becomes radiatively inefficient and no longer produces the copious UV required to illuminate the BLR. This effectively switches off the sources of external seed photons, so that the BHs produce BL Lac rather than FSRQ type jets. A few systems remain at high accretion rates - these typically host smaller BHs ($\sim10^7M_\odot$), which haven't yet used up their gas supplies. These correspond to Seyfert galaxies in the local Universe. According to our criteria ($\dot{m}>10^{-2}$), these BHs should host EC jets. However Fig.\ref{fig3}a shows that the Fermi visibility of jets from such small BHs is poor, so their contribution to the number of FSRQs at late times ($z<1$) is small. 

Whilst the predicted redshift distribution peaks at $1<z<2.5$, there is a tail out to high redshifts, with the most distant FSRQs being detected out to $z\sim5$. As redshift increases, the typical BH mass decreases. At $z=2$, the bulk of the accretion luminosity is produced by $10^8M_\odot$ BHs (Fig.\ref{fig3.5}). For $z>2$, the typical BH mass producing the bulk of the accretion luminosity drops below $10^8M_\odot$. Fig.\ref{fig3}a shows how sharply the Fermi visibility drops with mass, more than halving for a decade drop in mass from $10^9-10^8M_\odot$. Fig.\ref{fig3}b shows that for small BH mass ($\leqslant10^8M_\odot$), the increase in accretion rate at early times does not compensate for the drop in mass in terms of Fermi visibility (compare magenta and blue lines, Fig.\ref{fig3}b). The decreasing tail of the predicted redshift distribution from $2.5<z<5$ is therefore due to the decreasing number density of $10^{8-9}M_\odot$ BHs and the increasing reliance on strongly beamed sources ($\theta_{obs}\sim0$) to reach the Fermi flux limit. 

The total number of Fermi visible FSRQs predicted by our simulation is $\sim26000$, while the actual number of FSRQs detected in the Fermi 1 year catalogue is $\sim300$ (Abdo et al. 2010). Our simulation overpredicts the number of Fermi visible FSRQs by $\sim2$ orders of magnitude. This is one order of magnitude less than the 3 orders of magnitude discrepancy found in Paper 1 using the same method to predict the Fermi population of BL Lacs. Nevertheless, a 2 order of magnitude discrepancy is still too large to be explained by the sky coverage limit of Fermi ($|b|>10^\circ$ implying $80\%$ of the sky is included). In Fig.\ref{fig4}, we also show the observed redshift distribution of Fermi detected FSRQs from Shaw et al. (2012) (red line). The observed redshift distribution peaks at later times ($0.5<z<1.5$ rather than $1<z<2.5$), with no FSRQs detected in the 1LAT catalogue with $z>3.5$. Not only is the total number of FSRQs overpredicted, but the shape of the redshift distribution also does not match the observations. 

Fig.\ref{fig5}c\&d show the predicted mass and accretion rate distributions of Fermi visible FSRQs from the simulation, compared to the observed distributions (Fig.\ref{fig5}a\&b) measured by Shaw et al. (2012). The typical predicted FSRQ accretion rate is $-1<\log\dot{m}<0$, since Fermi visibility increases with accretion rate. Above Eddington, the increase in Fermi flux with $\dot{m}$ becomes progressively less (see Fig.\ref{fig2}a) and the number density of super Eddington sources of sufficient mass ($>10^8M_\odot$) drops off sharply, both of which result in the typical FSRQ accretion rate being just sub-Eddington. This is in rough agreement with the findings of Shaw et al. (2012), where most FSRQs are observed to have $-1<\log\dot{m}<0.5$. The typical predicted mass is $10^{8-8.5}M_\odot$, since these FSRQs are bright in the Fermi band and most numerous at $1<z<2$ where quasar activity peaks. The results of Shaw et al. (2012) show that the observed distribution is less sharply peaked and the peak extends to slightly higher mass ($10^{8-9}M_\odot$).

\subsection{Dependence on Black Hole Spin?}

\begin{figure} 
\centering
\begin{tabular}{l}
\leavevmode  
\epsfxsize=8cm \epsfbox{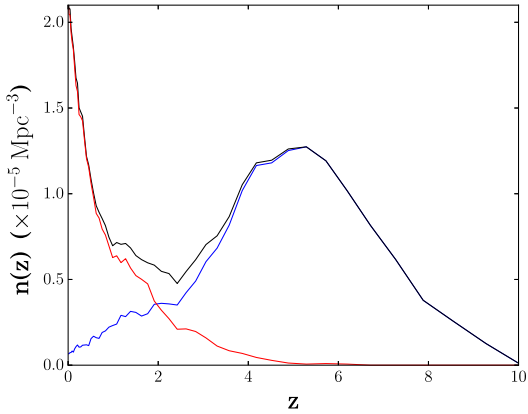} \\
\end{tabular}
\caption{Redshift distribution of high spin BHs ($a>0.8$) from the Millennium simulation, assuming accretion is chaotic at all redshifts. Red line shows BHs accreting with $\dot{m}<10^{-2}$ (corresponding to BL Lacs), blue line shows BHs accreting at $\dot{m}>10^{-2}$ (corresponding to FSRQs), black line shows total.}
\label{fig6}
\end{figure}

By assuming that every BH accreting with $\dot{m}>10^{-2}$ is capable
of producing a FSRQ jet, our simulation overpredicts the number of
Fermi detected FSRQs by two orders of magnitude. Clearly another
factor is reducing the number of FSRQs detected by Fermi. Paper 1
found that the number of Fermi detected BL Lacs was similarly
overpredicted (by 3 orders of magnitude) when the same technique was
applied to predict the observed population of BL Lacs (i.e. all BHs
accreting {\emph{below}} $10^{-2}$ produce BL Lac type jets). Paper 1
found that the observed numbers of BL Lacs, and their redshift
distribution, were much better reproduced assuming that only high spin
BHs ($a>0.8$) with $\dot{m}<10^{-2}$ produce BL Lac type jets. This
suggests BH spin might be important in the production of highly
relativistic $\Gamma=15$ jets in BL Lacs. Maraschi et al. (2012)
suggest that the efficiency of spin-powered jet production drops off
sharply below $0.8$, which provides additional support for an
effective spin threshold for relativistic jet production at
$a\sim0.8$ (although this is highly uncertain and there could be a more continuous
distribution of jet power with spin, e.g. Tchekhovskoy
et al. 2010). We investigate the effect of a sharp cut-off in jet
power with spin for the FSRQs in order to compare with the BL Lacs in
Paper 1. 

The cosmological simulations track the evolution of BH spins as well
as tracking their mass and accretion rate. BH spin is affected both by
accretion and by BH-BH coalescence following galaxy mergers (Volonteri
et al. 2005; 2007; Fanidakis et al. 2011; 2012). BH-BH mergers produce
highly spinning BHs, as the final merged BH is spun up by the angular
momentum of the orbiting merging BHs. The effect of accretion on BH
spin depends on the mode of accretion. For the case of prolonged
accretion, all the mass is accreted in a single event, with a single
angular momentum direction, which is sufficient to spin most BHs up to
maximum (Volonteri et al. 2005; 2007). If the accretion is chaotic,
with gas accreted in a series of smaller events that are randomly
aligned, the net angular momentum transfer to the BH is zero (King et
al. 2008). Chaotic accretion therefore results in predominantly low
spin BHs, with high spins being rare and only produced by BH-BH
mergers (Fanidakis et al. 2011; 2012).

Paper 1 found the chaotic accretion model was required to match the
population of Fermi detected BL Lacs. High spin in the chaotic
accretion model is rare, so requiring high spin reduces the predicted
number of BL Lacs, in better agreement with the observed
numbers. Requiring high spin also causes the predicted redshift
distribution to peak at later times (lower redshift) and increases the
typical predicted BL Lac mass (since gas poor mergers happen later and
produce the most massive BHs), in better agreement with
observations. Hence we choose the chaotic accretion model.

\begin{table*}
\begin{tabular}{lcccc}
\hline
Parameter & $<FSRQ>$ & $\gamma$NLS1 & Standard Scaling Prediction \\
\hline
$M$ ($M_\odot$) & $10^9$ & $1.5\times10^8$ & $1.5\times10^8$ \\
$\dot{m}$ ($L/L_{Edd}$) & $0.1$ & $0.5$ & $0.5$ \\
$R_{diss}$ ($\times10^{15}$cm) & $18.9$ & $6.75$ & $2.835$ \\
$P^\prime_{rel}$ ($\times10^{43} erg s^{-1}$) & $2.0$ & $2.3$ & $1.5$ \\
$B$ (G) & $2.6$ & $4.1$ & $15.0$ \\ 
\hline
\end{tabular}
\caption{Comparison of the observed jet parameters for the $\gamma$NLS1 PMN J0948+0022 (Abdo et al. 2009c) with those expected from scaling the mean FSRQ jet parameters from G10 according to standard jet scalings ($R_{diss}\propto M$, $P^\prime_{rel}\propto \dot{m}M$, $B\propto (\dot{m}/M)^{1/2}$).}
\label{table:gammarayNLS1comparison}
\end{table*}

The blue line in Fig.\ref{fig4} shows the predicted redshift
distribution of Fermi visible FSRQs after imposing a spin cut, so that
only BHs with $\dot{m}>10^{-2}$ and $a>a_{cut}$ produce a FSRQ type
jet. We find $a_{cut}\sim0.77$ is required to reproduce the observed
number of Fermi detected FSRQs. However, closer comparison of the
observed and predicted distributions (red and blue lines) shows that,
although the total number of FSRQs is better reproduced, the
simulation cannot reproduce the tail out to high redshifts
($>2$). Imposing a spin cut limits the maximum expected FSRQ redshift
to $\sim2.3$.

Fig.\ref{fig6} shows the number density of high spin BHs ($a>0.8$) as
a function of redshift from the Millennium Simulation (black
line). This has two peaks, one at $z=0$ and one at $z=5$. The red line
shows the number density of high spin BHs with low accretion rates
($\dot{m}<10^{-2}$). These are responsible for the peak at $z=0$. They
get their high spins from late gas poor mergers, so represent the most
massive BHs. This is the population of BHs responsible for the
production of BL Lac jets. The blue line shows the number density of
high spin BHs with high accretion rates ($\dot{m}>10^{-2}$). These are
responsible for the peak at $z=5$. They acquire their high spins
through early mergers of much smaller BHs. At early times the BHs
still have a plentiful gas supply (hence their high accretion rates)
and subsequent chaotic accretion gradually spins down the BHs, so that
the number density of high spin high accretion rate BHs drops with
decreasing redshift, not only because typical accretion rates drop,
but also because most BHs are losing their earlier high spins. If we
require high spin as well as high accretion rate to produce a FSRQ
type jet, then these are the BHs that should be responsible for
FSRQs. However, when we include a spin cut in our simulation, our
results cannot replicate the observed FSRQ population between
$2<z<3$. Fig.\ref{fig6} (blue line) shows that the number of high spin
high $\dot{m}$ BHs has dropped significantly by $2<z<3$. Many of the
high spin BHs that remain are still small ($10^{7-8}M_\odot$), because
if they had grown significantly since their last merger the process of
chaotic accretion would have reduced their spins. As a result, they
are not massive enough to be Fermi visible in our simulation. Yet the
observations show there are some relatively massive BHs with FSRQ jets
at these redshifts. Our simulation accounts for spin ups due to
mergers, so these BHs cannot have acquired their spins through
mergers, but must have maintained them whilst growing by accretion.

This suggests that early accretion may be more organised than late
accretion. Perhaps there is a transition from prolonged accretion to
more chaotic accretion as gas supplies diminish and redshift
decreases. In assuming a chaotic accretion mode throughout, we have
therefore underestimated the number of high spin BHs at higher
redshifts ($>2$), where accretion may be more ordered.

However, we know that, in order to reproduce the observed BL Lac
population, most BHs must have been reduced to low spins by $z=2$,
when accretion rates drastically drop and chaotic accretion takes
over. The problem then is, if BHs are not being spun down by chaotic
accretion at early times, what causes the BHs to lose their high spins
when their accretion rates drop at $z=2$?

There is one additional factor that affects BH spin, aside from
accretion mode and BH-BH mergers, that the simulation does not take
into account, and that is the jet itself. Powering a highly
relativistic jet should cause the BH to spin down (eg. Nemmen et
al. 2007, Tchekhovskoy 2011, Dotti et al. 2013). Perhaps at early
times ($z>2$), when accretion is more ordered, the angular momentum
the BH gains from the prolonged accretion flow balances the spin down
effect of the relativistic jet. Then, when the accretion rate drops at
$z\sim2$, the BH loses this input source of angular momentum and the
jet spins down the BH (which in turn switches off the highly
relativistic jet). A powerful jet can spin down a central black hole
in $3\times 10^8$ years (Tchekhovskoy \& McKinney 2012;
Tchekhovskoy \& Giannios 2015). As a result, most BHs at late times ($z<2$) are low
spin and the only high spin BHs are those which were subsequently spun
up in late gas poor BH-BH mergers.

Fig.\ref{fig5}e\&f show the predicted mass and accretion rate
distributions of Fermi visible FSRQs from our current simulation
including a spin cut. Both are in clear disagreement with the observed
distributions (Fig.\ref{fig5}a\&b). Imposing the spin cut has
preferentially selected for low accretion rate objects ($\dot{m}$
close to $10^{-2}$), which are typically higher mass
(Fig.\ref{fig3.5}). These objects are the high accretion rate end of
the BL Lac population --- either late mergers that were not gas poor,
or BHs around $z\sim2$ with decreasing accretion rates that are in the
process of transitioning to a radiatively inefficient accretion flow
and a BL Lac type jet. Our spin cut has not captured the population of
high accretion rate, relatively high mass objects ($10^8$-$10^9$) at
$1<z<3$ that make up the bulk of the observed FSRQ population. This
further emphasises that a more sophisticated model combining chaotic
and prolonged accretion episodes with jet spin down is required to
trace the evolution of BH spin beyond $z=2$, assuming highly
relativistic FSRQ jets really are a tracer of high spin objects.

\section{Caveats}

The Fermi band flux in FSRQs is dominated by Compton upscattering
of seed photons from the BLR. We have approximated the BLR as a
spherical shell of radius $R_{BLR}$ centred on the BH so that some
fraction of the seed photons come from directly ahead of the jet.
However, studies of line profiles have shown that the BLR geometry may
be more flattened, perhaps indicating an origin as a disc wind
(Kollatschny \& Zetzl 2013). Nonetheless, aberration would still mean
that these photons would appear close to face on in the jet frame for 
the bulk Lorentz factors assumed here, so this is unlikely to have
much of an effect. 

We have assumed that a fixed fraction of $L_d$ is reprocessed by the
BLR and torus. Again, this may not be the case. G10 find from
spectral fitting of a sample of FSRQs that the fraction does vary
slightly, although not by much, so this should not be important in our
statistical sample.

We have used the mean FSRQ spectrum of G10 as our model spectrum from
which to scale, however FSRQs are highly variable. During flaring
  the Fermi flux can increase by more than an order of magnitude. As a
  result, distant FSRQs that would not normally be detected may become
  visible. For example, Ghirlanda et al. (2011) note that some EGRET detected
  FSRQs were not visible in the first years of Fermi LAT observations
  despite its much better sensitivity, showing that the EGRET
  detection was only a short flaring episode. This would extend the
  tail of the redshift distribution out to higher redshifts than
  otherwise expected. In only modelling the typical FSRQ emission,
  such events have not been included in our simulation.

More importantly, we have also assumed that a FSRQ jet is produced for
the entire time a BH is accreting with $\dot{m}>10^{-2}$. If instead
the jet follows a duty cycle and is only produced for a fraction of
that time then this will reduce the number of Fermi detected FSRQs. We
overpredict the number of Fermi visible FSRQs by 2 orders of magnitude
assuming every BH with $\dot{m}>10^{-2}$ produces a FSRQ jet. If each
of these BHs only produces a FSRQ jet $100^{th}$ of the time of each
accretion episode, then we could match the observed numbers without
needing any limits on the spin of the BH. However, FSRQs should be the
aligned analogues of FRII sources (Padovani \& Urry 1992). The large
scale radio lobes of FRII sources indicate that the jet producing them
must be persistent, since the time taken to produce such large scale
structures is $\sim$Myr, probably similar to the time of each
accretion episode. One explanation may then be that the large scale
structure is produced by a persistent slower jet, while there is a
fast central spine with $\Gamma=13$, which appears as a FSRQ when
viewed head on and is intermittent.

\subsection{The Gamma Ray Loud NLS1 as a Test of Jet Scalings}\label{sec:GNLS1}

Our jet scalings reproduce the major trends seen in the FSRQ blazar
sequence (see Section 3.1 and 3.2). However, most of these objects are
high mass BHs. The small number of Gamma-ray loud NLS1s ($\gamma$NLS1s)
provide a unique opportunity to test the jet scalings on much smaller
mass ($10^{6-8}M_\odot$), high accretion rate AGN. These $\gamma$NLS1s are a
subset of the radio loud NLS1s (Komossa et al. 2006) which have been
detected by Fermi. They show Doppler boosted jet
emission with a weak synchrotron hump and strong IC emission so that
their SEDs appear like `mini FSRQs' (Abdo et al. 2009a; 2009b). The high
accretion rates of NLS1s ($\dot{m}\sim1$) mean that their accretion
flows are in the radiatively efficient regime, giving them a BLR
(albeit with relatively narrow broad lines), so they should correspond
to low mass FSRQs. In which case, we should be able to replicate their
spectra simply by turning down the mass in our mean FSRQ spectral
model.

PMN J0948+0022 was the first $\gamma$NLS1 to be discovered. A
multi-wavelength monitoring campaign was carried out on the source in
2009. Abdo et al. (2009c) subsequently fitted its broadband spectrum
with the jet model of Ghisellini \& Tavecchio (2009). In Table 1 we
show their derived values of $R_{diss}$, $P^\prime_{rel}$ and $B$. We also
show the expected values of $R_{diss}$, $P^\prime_{rel}$ and $B$ for this
source that result from scaling the mean FSRQ parameters of G10 as
$R\propto M$, $P^\prime_{rel}\propto \dot{m}M$ and $B\propto
(\dot{m}/M)^{1/2}$, according to the mass and accretion rate of PMN
J0948+0022 ($M=1.5\times10^8\,M_\odot$, $\dot{m}=0.5$). These scalings
correspond to assuming both the power injected into relativistic
electrons and the power in magnetic fields are a fixed fraction of the
accretion power, i.e. $P^\prime_{rel}\propto P^\prime_B\propto P_{acc} \propto
\dot{m}M$. In Table 1 we also list the mean FSRQ values for reference.

The values of $R_{diss}$ and $P^\prime_{rel}$ found from fitting the observed spectrum of PMN J0948+0022 are both slightly larger than expected from scaling from the mean FSRQ parameters. However, the biggest difference is in the magnetic field strength. Scaling the mean FSRQ magnetic field strength of $2.6$\,G as $B\propto (\dot{m}/M)^{1/2}$ implies PMN J0948+0022 should have a jet magnetic field of $15$\,G. In reality, the magnetic field required to fit its spectrum is much smaller ($4$\,G). This is larger than the mean FSRQ value, as expected for its smaller mass, but not nearly as large as the standard scalings predict. 

\begin{figure} 
\centering
\begin{tabular}{l}
\leavevmode  
\epsfxsize=8cm \epsfbox{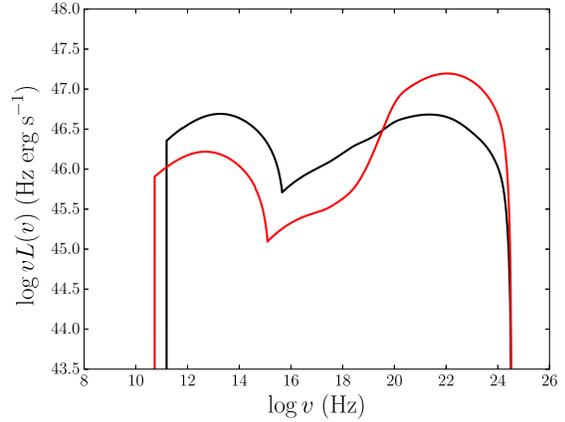} \\
\end{tabular}
\caption{Comparison of a $\gamma$NLS1 spectrum predicted using standard jet scalings with a spectrum using observationally constrained parameters. Black line shows expected jet spectrum for a BH with mass and accretion rate of the $\gamma$NLS1 PMN J0948+0022 ($M=1.5\times10^8\,M_\odot$, $\dot{m}=0.5$), from scaling the mean FSRQ spectrum of G10 according to standard jet scalings ($R_{diss}\propto M$, $P^\prime_{rel}\propto \dot{m}M$ and $B\propto (\dot{m}/M)^{1/2}$). Red line shows resulting spectrum replacing $R_{diss}$, $P^\prime_{rel}$ and $B$ with the observed values found by Abdo et al. (2009c) to fit PMN J0948+0022. See Table 1 for parameter values.}
\label{fig7}
\end{figure}

In Fig.\ref{fig7} we show the effect of this on the observed spectrum. The black line shows the spectrum produced taking the mean FSRQ parameters and scaling $R_{diss}$, $P^\prime_{rel}$ and $B$ to the mass and accretion rate of PMN J0948+0022 according to the standard scaling relations, i.e. standard scaling prediction from Table 1. The red line shows the same spectrum, but replacing $R_{diss}$, $P^\prime_{rel}$ and $B$ with the values found by Abdo et al. (2009c) from fitting the observed spectrum of PMN J0948+0022 (i.e. observed values in Table 1). The red line is not a fit to PMN J0948+0022, since we have kept the other parameters ($\phi$, $\Gamma$ etc.) the same as the mean FSRQ, in order to show just the effect of correcting $R_{diss}$, $P^\prime_{rel}$ and $B$ (although we note that the injected electron distribution parameters of PMN J0948+0022 are not very different to those of the mean FSRQ). It is clear that replacing the values predicted by the standard scalings with the observed values has a big effect on the shape of the spectrum. The high magnetic field predicted by the standard scaling relations causes synchrotron and SSC emission to dominate the black predicted spectrum, resulting in synchrotron and Compton peaks of similar luminosity. The resulting spectral shape is similar to that of a BL Lac, where the only source of emission is synchrotron and SSC emission. In contrast, when we use the observed values, where the magnetic field is much lower, the Compton emission is dominated by IC from external seed photons and the synchrotron emission is suppressed (red spectrum). As a result, the Compton peak is much brighter than the synchrotron peak --- a spectral shape typical of FSRQs. However, according to the standard scaling relations, $\gamma$NLS1s {\emph{shouldn't}} look like mini FSRQs --- SSC should dominate their Compton humps not IC. 

This leaves us with two potential scenarios. The first is that FSRQs simply do not follow standard scaling relations. This is surprising, given that these scaling relations are based on just two assumptions: that the size scales of the jet should scale with BH mass, and that the power injected into relativistic electrons and the power in magnetic fields is a fixed fraction of the accretion power. Fig.\ref{fig7} shows that replacing the parameters predicted by standard jet scalings with those observed for a $\gamma$NLS1 increases the Fermi flux ($\nu\sim10^{22}$) by half an order of magnitude. This suggests, if $\gamma$NLS1s really are low mass versions of FSRQs, then the low mass FSRQs in our simulation should be brighter and hence more visible than we have estimated. Consequently, our original predicted population, which already overestimates the observed population by 2 orders of magnitude, should be an underestimate. This only increases the need for some other factor, such as a limit on BH spin, to reduce the predicted numbers. 

The alternative scenario is that FSRQs do follow standard jet scalings and the masses of $\gamma$NLS1s have simply been underestimated. Several studies have suggested that $\gamma$NLS1 masses are systematically larger, and accretion rates correspondingly lower, than previously estimated (eg. Calderone et al. 2013; Baldi et al. 2016; D'Ammando et al. 2017). Given that $\gamma$NLS1s show strong Compton dominance, whereas standard jet scalings predict a low Compton dominance for typical NLS1 masses, our findings support this interpretation. 

\section{Comparing FSRQ and BL Lac Jets}

FSRQs and BL Lacs are typically understood as representing the two ends of the `Blazar sequence'. The transition from low power BL Lac to high power FSRQ can be understood in terms of increasing accretion rate onto the central BH. The dimmest BL Lacs, produced by the lowest accretion rate BHs ($\dot{m}<10^{-3}$), appear as high peaked BL Lacs (HBLs). Their low accretion rates mean lower magnetic fields and lower injected electron powers, which result in less cooling, so the synchrotron and SSC emission peak at high frequencies. As accretion rate increases, $B$ and $P^\prime_{rel}$ increase, the amount of cooling increases, so the electron distribution cools down to lower Lorentz factors and the observed synchrotron and SSC spectra peak at lower frequencies. Increasing $\dot{m}$, and the corresponding increase in $B$ and $P^\prime_{rel}$, switch the observed spectrum from a HBL to a low peaked BL Lac (LBL). As $\dot{m}$ becomes greater than $10^{-2}$, the accretion flow around the BH switches from a radiatively inefficient flow to a radiatively efficient UV accretion disc, effectively turning on external sources of seed photons, and the jet stops being a BL Lac and appears as a FSRQ. In this picture, the jet is the same in both cases, the only difference is in the power input ($B$ and $P^\prime_{rel}$) and the presence or absence of external seed photons, both of which are linked by a dependence on the accretion rate. However, in reality this is not the case. There are further differences between the two types of jet. 

Comparison of the mean injected electron distribution parameters found by G10 for FSRQs and BL Lacs shows that $\gamma_{max}$ and $\gamma_b$ are much larger for BL Lacs ($\gamma_{max}\sim10^5$ for BL Lacs compared to $10^3$ for FSRQs and $\gamma_b\sim10^4$ compared to $10^2$). The difference between BL Lacs and FSRQs is not simply that the electrons have a different seed photon field to cool off. The accelerated electron distribution is intrinsically different in FSRQs compared to BL Lacs. This suggests there is some difference in the way electrons are accelerated, presumably by shocks, in FSRQ jets compared to BL Lacs. 

A more fundamental difference is in jet opening angle ($\phi$). Here and in Paper 1 we have used $\phi=0.1$, which is typically assumed for calculating blazar spectra. However, Krause et al. (2012) have shown that $\phi$ should be larger for BL Lac jets. They find from hydrodynamic simulations that jet opening angle sets the large scale morphology of the jet, with FRII jets (corresponding to misaligned FSRQs) being produced for opening angles $<24^\circ$ ($=0.4$\,rad) and FRI morphologies (corresponding to misaligned BL Lacs) being produced for larger opening angles. Since $\phi$ relates $Z_{diss}$ and $R_{diss}$, this means that the same size emission region will be located at smaller $Z_{diss}$ for a larger opening angle. Since the calculation of BL Lac spectra does not involve any external seed photons, the only change as a result of accounting for a larger opening angle in BL Lacs will be that the $Z_{diss}$ derived from fitting a given BL Lac spectrum will be smaller. 

A related factor is that the mean BL Lac BLF is slightly larger than the mean FSRQ BLF (15 compared to 13, G10). The BLF of the jet should influence where the dissipation region is, if it corresponds to a standing shock at the base of the jet. For larger $\Gamma$ material will travel further before shocking. The discontinuity in both opening angle and BLF suggests that $R_{diss}$ and $Z_{diss}$ should not scale continuously between FSRQs and BL Lacs.

\section{Conclusions}

We have combined models of FSRQ spectra together with prescriptions
for how they should scale with mass and accretion rate and the number
densities of BHs from cosmological simulations to predict the number
of FSRQs that should be observed by Fermi. If we assume all BHs
accreting with $\dot{m}>10^{-2}$ produce a FSRQ jet, our simulation
overpredicts the number of Fermi detected FSRQs by two orders of
magnitude. If we restrict the production of FSRQ jets to high spin BHs
($a>0.77$), we can reproduce the observed numbers. However, our
predicted redshift distribution does not extend to as high redshift
($2<z<3$) as the observed redshift distribution and we cannot match
the observed mass and accretion rate distributions of FSRQs.

We suggest this may reflect a limitation in the ability of the
cosmological simulations to track the evolution of black hole spin. If
production of FSRQ jets really does require a high spin BH, our
simulations suggest there should be more high mass high spin BHs at
high redshift ($\sim2-3$) than a solely chaotic accretion model
predicts. In a chaotic accretion model, high spin is rare and only
achieved through BH-BH mergers. Therefore the number density of high
spin BHs peaks at $z\sim5$ (corresponding to the first mergers of high
accretion rate BHs) and $z\sim0$ (corresponding to late gas poor
mergers of the most massive, lowest accretion rate BHs). Our
simulations lack high mass high spin BHs at redshift $z=2-3$ because
chaotic accretion spins down the BHs as they grow. Yet the observed
redshift distribution of FSRQs requires moderately massive ($10^8
M_\odot$) BHs at $z=2-3$ that are high spin, assuming the FSRQ
population is a tracer of high accretion rate BHs with high spin. This
means these BHs must have maintained high spins while they were
accreting, suggesting a trend from chaotic accretion (which spins down
BHs) towards more prolonged accretion (which can spin up/maintain BH
spin) at high redshifts. This suggests accretion must have been more
ordered in the early Universe, at least for some objects
(Dotti et al.
2013; Dubois et al. 2014).

However, Paper 1 showed that, for $z<2$, a chaotic accretion model is
required to reproduce the observed population of BL Lacs. This means
that, by $1.5<z<2$, most BHs must have been reduced to low spin, so
that the only high spin BHs in the local Universe are the most massive
BHs, which have undergone late gas poor mergers. The problem then is
how to explain the sudden reduction in BH spin in those objects that
were previously maintaining their high spin via prolonged
accretion. This loss of spin may be caused by the jet itself, since a Blandford-Zdnajek jet is powered by the spin energy. The
  resultant timescale for spin-down depends on the balance of this
  extraction of spin energy with spin up from accretion of the same
  material which brings in the magnetic field to power the jet (Wilson
  \& Colbert 1995; Moderski \& Sikora 1996). While these objects are
highly accreting at $z>2$, the angular momentum the BH gains from the
prolonged accretion flow balances the spin down effect of the
relativistic FSRQ jet. Then, when the accretion rate begins to drop at
$z\sim2$, the BH loses this input source of angular momentum and the
jet spins down the BH, which in turn switches off the highly
relativistic jet. A powerful jet can spin down a central black hole
in $3\times 10^8$ years (Tchekhovskoy \& McKinney 2012;
Tchekhovskoy \& Giannios 2015), which is sufficient to reduce most previously high spin BHs to low spin at late times ($z < 2$).

An important additional factor in our simulations are the scaling
relations we use to predict the spectra from FSRQs of different masses
and accretion rates. We test these by comparing our scaled models with
gamma-ray loud NLS1s ($\gamma$NLS1s), which should be scaled down
versions of the more massive FSRQs. We find that standard scaling
relations (allowing all sizescales to scale with mass and assuming the
power injected into relativistic electrons and magnetic fields is a
constant fraction of the accretion power) predict $\gamma$NLS1s
spectra should be much less Compton dominant than observed. On face
value, this suggests that, for some reason, FSRQ jets do not follow
standard jet scaling relations. However, an alternative explanation
may be that $\gamma$NLS1 masses are not as low as previously
estimated. In which case, the high Compton dominance of $\gamma$NLS1
spectra supports the suggestion of other recent studies (eg. Calderone
et al. 2013; Baldi et al. 2016; D'Ammando et al. 2017) that
$\gamma$NLS1 masses have been systematically underestimated.

It is clear that not every BH accreting with $\dot{m}>10^{-2}$ can
produce a FSRQ jet, just as not every BH accreting with
$\dot{m}<10^{-2}$ can produce a BL Lac jet. Restricting highly
relativistic jet production to high spin BHs produces a good match to
the observed population of BL Lacs (Paper 1), so it is likely that
high spin may be similarly important for FSRQs, whose jets appear to
be the high accretion rate analogues of BL Lacs. In which case,
combining the observed redshift distributions of BL Lacs and FSRQs
should allow us to trace the population of high spin BHs as a function
of redshift. This provides a powerful observational constraint to test
whether new models combining chaotic and prolonged accretion with jet
spin-down really can track the evolution of BH spin across cosmic
time.

\section{Acknowledgements}

We thank Nikos Fanidakis and the Millennium simulation for use of their data. This work has made use of Ned Wright's Cosmology Calculator (Wright 2006). EG acknowledges funding from the UK STFC.

\appendix 
\section{}

The jet emission model is publicly available as the {\sc{xspec}} local model {\sc{jet}}. This is the single zone, leptonic, relativistic jet model developed by Ghisellini \& Tavecchio (2009), as used in Ghisellini et al. (2010), as coded up by Gardner \& Done for this work. Please reference all three papers if you use this model in {\sc{xspec}}. It can be used in conjunction with {\sc{optxagnf}}, which models the emission from the accretion flow. In which case, the first three parameters (BH mass, comoving distance and accretion rate) can be tied together. 

Table A1 lists the model parameters. The first three set the parameters of the BH and the distance. Parameters 4-7 set the physical parameters of the jet: inclination to the line of sight, BLF, jet opening angle and distance of the emission region from the BH. When combined together the last two of these set the radius of the emission region, since the code assumes a conical jet. Parameters 8 and 9 set the jet magnetic field and the power injected into relativistic electrons. Parameters 10-14 determine the shape of the injected electron distribution, and parameter 15 sets the redshift. {\sc{xspec}} requires a 16th normalization parameter, which must be fixed at unity. 

The code can be used to model both FSRQs and BL Lacs. If $\log\dot{m}<-2$ (parameter 3), the code assumes the accretion flow regime is radiatively inefficient and there is no UV bright accretion disc. In this case, the external seed photon energy density is set to zero and the model calculates only synchrotron and SSC emission, producing a BL Lac type jet. If $\log\dot{m}\geqslant -2$, then the model assumes a radiatively efficient accretion disc is present and it includes IC emission from external sources of seed photons by assuming the radiatively efficient disc illuminates the BLR and torus, both of which reprocess a fraction of the disc emission. In this case, the code calculates the energy density of seed photons from direct disc and coronal emission, BLR emission, reflection of coronal X-rays off the BLR and emission from the torus, following the method of Ghisellini \& Tavecchio (2009). 

The code prints to screen which type of jet is calculated (SSC or SSC+IC), along with the logarithm of the power in radiation, magnetic fields, electrons, protons and total jet power ($P_r$, $P_B$, $P_e$, $P_p$ and $P_j$), where all five powers are in the observer's frame. For SSC+IC jets, the code prints to screen $R_{BLR}$ and $R_{IR}$ and flags if $Z_{diss}>R_{BLR}$ and $Z_{diss}>R_{IR}$. 

The various spectral components (synchrotron, SSC, EC disc, EC X-ray corona, EC BLR, EC X-ray reflection from BLR and EC torus) can be written out individually, by editing the code and uncommenting the six lines beginning {\sc{write(2,*)}}. This produces 
file fort.2 in the directory where {\sc{xspec}} is being run. Uncommenting the lines beginning {\sc{write(3,*)}} writes out the energy density of seed photons in the jet frame ($U^\prime(\nu^\prime)$) in file fort.3, writing out in order: disc, coronal, BLR, reflected coronal X-rays and torus seed photon energy densities (if including IC) and, lastly, the energy density of synchrotron seed photons. Uncommenting the line beginning {\sc{write(908,*)}} will write the steady state electron distribution ($N(\gamma)$) to fort.908. 

Since the inclination of the jet is a parameter, the code can in practice be used to model any jet, not just highly aligned blazars. However, the code assumes a single emission zone, so it is best suited for modelling the high energy jet base emission. Although the {\sc{fortran}} subroutine can be easily modified to be called multiple times with increasing emission region size to model more extended structures.

\subsection{Jet Emission Calculation}

The emission comes from a single spherical zone of radius $R_{diss}$. We assume the jet has a constant opening angle ($\phi$), such that the distance of the emission region from the central BH ($Z_{diss}=z_{diss}R_g$) is related to the radius of the emission region by: $R_{diss}=\phi Z_{diss}$. We assume material in the jet moves at a constant bulk Lorentz factor ($\Gamma$) and that some fraction of the transported electrons are accelerated into a power law distribution between minimum and maximum Lorentz factors $\gamma_{min}$ and $\gamma_{max}$, of the form: 

\begin{multline}
Q(\gamma)=Q_0 \frac {\left(\frac{\gamma}{\gamma_b}\right)^{-s_1} }
{ \left[ 1+\left(\frac{\gamma}{\gamma_b}\right)^{-s_1+s_2}\right] }=Q_0q(\gamma) \\
\mbox{ for } \gamma_{min}<\gamma<\gamma_{max}
\end{multline}

$\gamma_b$ is the Lorentz factor at which the electron distribution changes in slope from $s_1$ to $s_2$. We calculate the normalisation $Q_0$ from the power injected into the accelerated electrons ($P^\prime_{rel}$):

\begin{equation}
P^\prime_{rel} = \frac{4\pi}{3} R_{diss}^3m_ec^2Q_0\int_{\gamma_{min}}^{\gamma_{max}}\gamma q(\gamma)d\gamma
\end{equation}

We calculate $\gamma_{cool}$ after a light crossing time $t_{cross}=R_{diss}/c=\gamma_{cool}/\dot{\gamma}_{cool}$, as: 

\begin{equation}
\gamma_{cool}=\frac{3m_ec^2}{4\sigma_TRU^\prime_{seed}}
\end{equation}

\noindent where $U^\prime_{seed}=U^\prime_B+U^\prime_{sync}+U^\prime_{ex}$ is the sum of the energy density in magnetic fields, synchrotron emission and external emission which provides the seed photons for cooling. 

We solve the continuity equation to find the self-consistent steady state electron distribution: 

\begin{equation}
\begin{split}
N(\gamma,t_{cross}) &= Kn(\gamma) \\
 &=
  \begin{cases} 
    AQ_0q(\gamma) \\ &\mbox{ for } \gamma_{min} < \gamma < \gamma_{cool} \\
    \frac{3m_ec^2}{4\sigma_TcU^\prime_{seed}}\frac{Q_0}{\gamma^2}\int_{\gamma}^{\gamma_{max}}q(\gamma)d\gamma \\ &\mbox{ for } \gamma_{cool} < \gamma <\gamma_{max} 
   \end{cases}
\end{split}
\end{equation}

\noindent where $A$ is found by matching at $\gamma_{cool}$. 

We use the delta function approximation and calculate the synchrotron emissivity as: 

\begin{equation}
j^\prime_{sync}(\nu^\prime)=\frac{\sigma_Tc}{6\pi\nu^\prime_B}U^\prime_B\gamma N(\gamma)
\end{equation}

\noindent where the electron Lorentz factor and synchrotron photon frequency are related by $\gamma=\sqrt{3\nu^\prime/4\nu^\prime_B}$ and we calculate the synchrotron self-absorption frequency ($\nu^\prime_{ssa}$) as given by (Ghisellini et al. 1985):

\begin{equation}
\nu^\prime_{ssa}=\left(4.62\times10^{14}KB^{2.5}\frac{R_{diss}}{0.7}\right)^{2/7}
\end{equation}

We calculate Compton emission including the Klein-Nishina cross section using the delta approximation:

\begin{equation}
j^\prime_{comp}(\nu^\prime)=\frac{\sigma_Tc}{6\pi}\int^{\gamma_{max}}_{\gamma_{min}}\int^{\nu^\prime_{seed,max}}_{\nu^\prime_{seed,min}}\frac{U^\prime_{seed}(\nu^\prime_{seed})}{\nu^\prime_{seed}}\gamma N(\gamma)d\nu^\prime_{seed}d\gamma
\end{equation}

\noindent where the electron Lorentz factor and Compton photon frequency are related by $\gamma=\sqrt{3\nu^\prime/4\nu^\prime_{seed}}$. 

Bulk motion of the jet boosts and blue shifts the emission. We calculate the observed flux as: 

\begin{equation}
F(\nu^\prime\delta/(1+z))=\frac{(j^\prime_{sync}(\nu^\prime)+j^\prime_{comp}(\nu^\prime))}{R_{co}^2}\frac{4\pi}{3} R_{diss}^3\delta^3
\end{equation}

\noindent where $\delta=(\Gamma-\cos{\theta}\sqrt{\Gamma^2-1})^{-1}$ is the Doppler factor and $R_{co}$ is the comoving distance to the object at redshift $z$. 

We neglect photon-photon pair production. However, the code calculates the source compactness and flags a warning if $l'\geqslant30$. This corresponds to $\tau_{\gamma\gamma}\sim1$, i.e. when the source starts to become optically thick to photon-photon pair production and this effect becomes important. For most blazar jets the compactness is typically $<3$.

\subsection{External Seed Photons}

If $\log\dot{m}\geqslant -2$, the model assumes a radiatively efficient accretion disc is present and includes IC emission from external sources of seed photons, calculating the energy density of seed photons following the method of Ghisellini \& Tavecchio (2009). The model includes direct disc and coronal emission, BLR emission, reflection of coronal X-rays off the BLR and emission from the torus. 

The accretion disc luminosity ($L_d$) is calculated from $M$ and $\dot{m}$ (parameters 1 and 3). Each annulus of the disc is seen at a different angle with respect to the jet emission region so receives a difference amount of Doppler deboosting ($b_d$). We approximate the energy density of disc seed photons from each annulus in the jet frame as:

\begin{equation}
U^\prime_d(\nu^\prime)=\frac{4\pi h b_d}{c^3}\frac{(\nu^\prime/b_d)^3}{\exp\left[\frac{h\nu^\prime /b_d}{kT}\right]-1}d\mu_d
\end{equation}

\noindent where $b_d=\Gamma(1-\beta\mu_d)$, $\mu_d=\cos\eta$ and $\eta$ is the angle of the annulus with respect to the jet axis. $\mu_d$ therefore varies between $\mu_{max}=1$, for the innermost radii which are directly behind the jet and experience most deboosting, to $\mu_{min}=\left[1+(R_{d,max}/Z_{diss})^2\right]^{-1/2}$ for the outermost radius $R_{d,max}=1000R_g$. We calculate the temperature of each disc radius from the mass and accretion rate input in parameters 1 and 3. 

We assume the luminosity of coronal X-rays is $L_X=f_XL_{d}=0.1L_{d}$ and the corona extends to $R_X=60R_g$. Its emission is therefore deboosted by a factor $b_X=\Gamma(1-\beta\mu_X)$, where $\mu_X=\left[1+(R_{X}/Z_{diss})^2\right]^{-1/2}$. The total energy density of coronal seed photons in the jet frame is therefore: 

\begin{equation}
U^\prime_X = \frac{f_XL_d\Gamma^2}{\pi R_X^2 c}\left[1-\mu_X-\beta(1-\mu_X^2)+\frac{\beta^2}{3}(1-\mu_X^3)\right]
\end{equation}

We assume the spectrum of this emission is a cut off power law starting from $b_X\nu_{d,peak}$ in the jet frame, where $\nu_{d,peak}=4kT_{max}/h$ is the frequency at which the unboosted disc spectrum peaks. We assume the power law cut off $v_{c}=150\times10^3e/h$, so that: 

\begin{equation}
U^\prime_X(\nu') \propto \nu'^{-\alpha_X}\exp\left[-\frac{\nu'}{b_X\nu_{c}}\right]
\end{equation}

\noindent where $\alpha=1$. 

We assume a fraction $f_{BLR}=0.1$ of the disc luminosity is reprocessed by the BLR. This emission takes the form of a BB centred on the frequency of the Lyman $\alpha$ line ($\nu_{Ly \alpha}=\frac{c}{4(1216\times10^{-8})}$), so that: 

\begin{equation}
U^\prime_{BLR}(\nu') \propto \frac{\nu'^3}{\exp\left[\frac{\nu'}{b_{BLR}\nu_{Ly \alpha}}\right]-1}
\end{equation}

The total energy density in the jet frame ($U^\prime_{BLR}$) and boosting factor ($b_{BLR}$) depend on the radius of the BLR ($R_{BLR}$) compared to $Z_{diss}$. The radius of the BLR scales with $L_d$ as: 

\begin{equation}
R_{BLR} = 10^{17}(\frac{L_d}{10^{45} erg\,s^{-1}})^{1/2} cm 
\end{equation}

If $Z_{diss}<R_{BLR}$:

\begin{equation}
U^\prime_{BLR} = \frac{17\Gamma^2}{12}\frac{f_{BLR}L_d}{4\pi c R_{BLR}^2}
\end{equation}
\begin{equation}
b_{BLR}= \Gamma
\end{equation}

If $Z_{diss}>3R_{BLR}$:

\begin{equation}
U^\prime_{BLR} = \frac{f_{BLR}L_d}{4\pi c R_{BLR}^2}\frac{\Gamma^2}{3\beta}\left[2(1-\beta\mu_1)^3-(1-\beta\mu_2)^3-(1-\beta)^3\right]
\end{equation}
\begin{equation}
\mu_1=\left[1+(R_{BLR}/Z_{diss})^2\right]^{-1/2}
\end{equation}
\begin{equation}
\mu_2=\left[1-(R_{BLR}/Z_{diss})^2\right]^{1/2}
\end{equation}
\begin{equation}
b_{BLR} = \Gamma(1-\beta\mu_1)
\end{equation}

If $R_{BLR}\leqslant Z_{diss} \leqslant 3R_{BLR}$, we use a power law interpolation between the two regimes for $U^\prime_{BLR}$ and use $b_{BLR}=\Gamma(1-\beta\mu_1)$ for $Z_{diss}=3R_{BLR}$. 

We assume a fraction $f_{XBLR}=0.01$ of the coronal X-rays are reflected by the BLR clouds. We assume the reflected emission has the same cut off power law shape as the direct coronal emission. Both $b_{XBLR}$ and $U^\prime_{XBLR}$ vary as $b_{BLR}$ and $U^\prime_{BLR}$, with $f_{BLR}L_d$ replaced with $f_{XBLR}f_XL_d$. 

We assume a fraction $f_{IR}=0.3$ of the disc luminosity is reprocessed by the torus. This emission takes the form of a BB at $\sim370$K (i.e. $\nu_{IR}=370k/h$), so that: 

\begin{equation}
U^\prime_{IR}(\nu') \propto \frac{\nu'^3}{\exp\left[\frac{\nu'}{b_{IR}\nu_{IR}}\right]-1}
\end{equation}

As in the case of the BLR seed photons, $U^\prime_{IR}$ and $b_{IR}$ depend on the radius of the torus ($R_{IR}$) compared to $Z_{diss}$. $R_{IR}$ scales with $L_d$ as:

\begin{equation}
R_{IR} = 2.5\times10^{18}(\frac{L_d}{10^{45} erg\,s^{-1}})^{1/2} cm
\end{equation}

\noindent and again we consider three regimes. If $Z_{diss}<R_{IR}$:

\begin{equation}
U^\prime_{IR} = \frac{f_{IR}L_d\Gamma^2}{4\pi c R_{IR}^2}
\end{equation}

If $Z_{diss}>3R_{IR}$:

\begin{equation}
U^\prime_{IR} = \frac{f_{IR}L_d}{4\pi c R_{IR}^2}\frac{\Gamma^2}{3\beta}\left[2(1-\beta\mu_1)^3-(1-\beta\mu_2)^3-(1-\beta)^3\right]
\end{equation}
\begin{equation}
\mu_1=\left[1+(R_{IR}/Z_{diss})^2\right]^{-1/2}
\end{equation}
\begin{equation}
\mu_2=\left[1-(R_{IR}/Z_{diss})^2\right]^{1/2}
\end{equation}

If $R_{IR}\leqslant Z_{diss} \leqslant 3R_{IR}$, we use a power law interpolation between the two regimes for $U^\prime_{IR}$. In all three cases we use $b_{IR}=\Gamma(1-\beta\mu_{IR})$, where $\mu_{IR}=\cos(\arctan(R_{IR}/Z_{diss}))$.

\begin{table}
\begin{tabular}{lll}
\hline
 & Parameter & Description \\
\hline
1 & $M$ & BH mass in solar masses \\
2 & $R_{co}$ & Comoving distance in Mpc \\
3 & $\log\dot{m}$ & Logarithm of mass accretion rate in units of $L/L_{Edd}$ \\
 & & (if $\log\dot{m}<-2$, code does SSC with no external seed photons) \\
4 & $\theta_{obs}$ & Inclination of jet axis from line of sight in degrees \\
5 & $\Gamma$ & Jet bulk Lorentz factor \\
6 & $\phi$ & Jet opening angle in radians \\
7 & $z_{diss}$ & Distance of dissipation region from BH in Rg \\
  & & (radius of dissipation region, $r_{diss} = \phi z_{diss}$) \\
8 & $B$ & Magnetic field in Gauss \\
9 & $\log P^\prime_{rel}$ & Power injected into relativistic electrons in the jet frame in $erg\,s^{-1}$ \\
10 & $\gamma_{min}$ & Minimum Lorentz factor of injected electron distribution \\
11 & $\gamma_b$ & Lorentz factor of break in injected electron distribution \\
12 & $\gamma_{max}$ & Maximum Lorentz factor of injected electron distribution \\
13 & $s_1$ & Index of injected electron distribution below the break \\
14 & $s_2$ & Index of injected electron distribution above the break \\
15 & $z$ & Redshift \\
16 & norm & Normalisation - must be fixed at unity \\
\hline
\end{tabular}
\caption{Summary of the {\sc{jet}} model parameters.}
\label{table:modelparameters}
\end{table}

\label{lastpage}

\end{document}